\documentclass[aps,prd,eqsecnum,12pt,showpacs,preprintnumbers,nofootinbib]{revtex4-2}
\usepackage{geometry}                
\usepackage[pdftex]{graphicx}
\usepackage[font=small,format=plain,labelfont=bf,up,textfont=normal,up,justification=justified,singlelinecheck=false]{caption}
\usepackage{caption,subcaption}
\usepackage{float,placeins,relsize}

\usepackage{float}
\usepackage{amssymb,amsmath,amsfonts,amsopn,bm,dsfont,mathrsfs}
\usepackage{mathtools}

\usepackage[mathscr]{euscript}
\DeclareMathAlphabet{\mathpzc}{OT1}{pzc}{m}{it}

\newcommand{\mf}{\mathlarger{\mathpzc{f}}}
\newcommand{\mh}{\mathlarger{\mathpzc{h}}}
\newcommand{\zp}{\mathlarger{\mathpzc{P}}}

\newcommand{\sA}{\mathscr{A}}
\newcommand{\sB}{\mathscr{B}}
\newcommand{\sC}{\mathscr{C}}
\newcommand{\sD}{\mathscr{D}}

\usepackage[colorlinks,linkcolor=blue, pdfstartview=FitH]{hyperref}
\usepackage{enumitem}
\usepackage{txfonts}
\usepackage{wrapfig}
\usepackage[compact,small]{titlesec}

\usepackage{amsbsy}

\hypersetup{linkcolor=blue,citecolor=blue,filecolor=black,urlcolor=blue}

\IfFileExists{dsfont.sty}
	{\usepackage{dsfont}
         \let\mathbb=\mathds
         \newcommand{\id}{\mathds{1}}}
	{\typeout{Package dsfont.sty was not found, using alternative macros.}
         \let\mathds=\mathbb
         \newcommand{\id}{\mbox{1 \kern-.59em \textrm{l}}}}

\renewcommand\a{\alpha}
\renewcommand\b{\beta}
\renewcommand\d{\delta}
\newcommand\ka{\kappa}
\renewcommand\l{\lambda}
\renewcommand\r{\rho}

\renewcommand\t{\tau}

\renewcommand\j{\psi}
\renewcommand\th{\theta}

\newcommand\e{\epsilon}
\newcommand\g{\gamma}
\newcommand\z{\zeta}
\newcommand\m{\mu}
\newcommand\n{\nu}

\newcommand\p{\pi}
\newcommand\h{\eta}
\newcommand\s{\sigma}
\newcommand\f{\phi}
\newcommand\w{\omega}
\newcommand\ve{\varepsilon}
\newcommand\vk{\varkappa}
\newcommand\vl{\mathlarger{\varrho}}

\newcommand\vf{\raisebox{2pt}{$\varphi$}}

\newcommand\La{\Lambda}
\renewcommand\P{\Pi}
\renewcommand\S{\Sigma}

\newcommand\D{\Delta}
\newcommand\G{\Gamma}
\newcommand\F{\Phi}

\newcommand\T{\Theta}
\newcommand\Y{\Upsilon}
\newcommand\W{\Omega}

\newcommand{\lag}{\langle}
\newcommand{\rag}{\rangle}

\newcommand{\cA}{{\cal A}}

\newcommand{\cD} {{\mathcal D}}
\newcommand{\tF} {{\widetilde F}}

\newcommand{\cL}{{\cal L}}

\newcommand{\cO}{{\cal O}}
\newcommand{\cS}{{\cal S}}

\newcommand{\cR}{{\cal R}}

\newcommand{\pa}{\partial}

\newcommand{\nn}{\nonumber \\}
\newcommand{\na}{\nabla}

\newcommand{\sdfrac}[2]{\mbox{\small$\displaystyle\frac{\raisebox{-1.8pt}{${#1}$}}{\raisebox{1pt}{${#2}$}}$}}

\newcommand{\rH}{r_{\!_H}}
\newcommand{\rM}{r_{\!_M}}
\newcommand{\br}{\bar r}
\newcommand{\bea}{\begin{eqnarray}}
\newcommand{\eea}{\end{eqnarray}}
\newcommand{\be}{\begin{equation}}
\newcommand{\ee}{\end{equation}}
\newcommand{\bes}{\begin{subequations}}
\newcommand{\ees}{\end{subequations}}

\def\nbox#1#2{\vcenter{\hrule \hbox{\vrule height#2in
\kern#1in \vrule} \hrule}}
\def\sq{\,\raise1pt\hbox{$\nbox{.10}{.10}$}\,}
\def\sqb{\,\raise.5pt\hbox{$\overline{\nbox{.09}{.09}}$}\,}

\newcommand{\sumi}{\raisebox{-1.5ex}{$\stackrel{\textstyle\sum}{\scriptstyle i}$}}
\newcommand{\sqrfr}[2]{\sqrt{\frac{\raisebox{-2pt}{$#1$}}{\raisebox{1pt}{\!$#2$}}}}

\usepackage[T1]{fontenc} 
\usepackage[titletoc]{appendix}
\usepackage{setspace}

\allowdisplaybreaks[4]

\begin{document}
\pagestyle{plain}

\title{Gravitational Vacuum Condensate Stars in the Effective Theory of Gravity}

\author{Emil Mottola}
\email{mottola.emil@gmail.com, emottola@unm.edu}
\affiliation{Department of Physics and Astronomy, University of New Mexico\\
Albuquerque NM 87131\\}

\begin{abstract}
The low energy effective theory of gravity comprises two elements of quantum theory joined to classical general relativity. 
The first is the quantum conformal anomaly, which is responsible for macroscopic correlations on light cones and a stress tensor 
that can strongly modify the classical geometry at black hole horizons. The second is the formulation of vacuum energy as
$\La_{\rm eff}\!\propto\! F^2$ in terms of an exact $4$-form abelian gauge field strength $F\!=\!dA$. When $A$ is identified 
with the Chern-Simons $3$-form of the Euler class, defined in terms of the spin connection, a $J\cdot A$ interaction is generated 
by the conformal anomaly of massless fermions. Due to the extreme blueshifting of local frequencies in the near-horizon region of 
a `black hole,' the lightest fermions of the Standard Model can be treated as massless there, contributing to the anomaly and 
providing a $3$-current source $J$ for the `Maxwell' equation~$d\ast F = \ast J$. In this phase boundary region, torsion is activated, 
and $F$ can change rapidly. The Schwarzschild black hole horizon is thereby replaced by a surface, with a positive surface 
tension and $\mathbb{R}\otimes \mathbb{S}^2$ worldtube topology, separating regions of differing vacuum energy. The result 
is a gravitational vacuum condensate star, a cold, compact, horizonless object with a $p_{_V}\!=\! - \r_{_V}$ zero 
entropy, non-singular de Sitter interior and thin quantum phase boundary layer at the Schwarzschild radius $2GM/c^2$. 
\end{abstract} 

\maketitle
\vfil
\eject

\tableofcontents
\pagenumbering{arabic}
\pagestyle{plain}
\vfil\break

\section{Introduction: The Question of the Final State of Gravitational Collapse}

Classical black holes (BHs) are vacuum solutions of Einstein's eqs.,\,whose characteristic feature is an event horizon, the
mathematical boundary separating the BH exterior from its interior, at which light hovers indefinitely, unable to
escape. If continued through the horizon, BH interior solutions harbor spacetime singularities of infinite density that are causally 
disconected from external observers~\cite{HawkEllis:1973}. 

The presumption that such a singular end state can result from realistic collapsing matter depends upon assumptions for the
matter stress-energy tensor source $T^\m_{\ \n}$, typically expressed by energy conditions. The strong 
energy condition ($\r + 3p \ge 0$) is needed to prove the convergence of timelike geodesics in the Hawking-Penrose
singularity theorem~\cite{HawkPen:1970,HawkEllis:1973}. A reasonable assumption for classical matter 
or radiation, this strong energy condition is violated both by quantum effects~\cite{BarVisser:2002}, and by the 
cosmological dark energy assumed responsible for the accelerating expansion of the universe~\cite{RiessSN:1998, PerlmutterSN:1999}.
The crucial property of vacuum dark energy is that it is characterized by negative pressure $p=\!-\r$, hence $\r + 3p < 0$, which leads 
to the defocusing rather than focusing of timelike geodesics, in which case gravity acts as a effectively repulsive force, counteracting 
collapse to a singularity. 

The somehat weaker null energy condition $\r + p \ge 0$ leads also to a singularity, if the existence 
of a closed trapped surface is assumed~\cite{Penrose:1965}. This assumption relies upon an essentially classical view of matter 
as uncorrelated pointlike particles, which fall freely through the horizon, with vanishing or negligibly small stress-energy 
tensor $T^\m_{\ \n}\! \approx \!0$ there. However Standard Model (SM) matter is described by quantum field theory (QFT) 
in which $T^\m_{\ \n}$ becomes a quantum operator $\hat  T^\m_{\ \n}$.  The renormalized expectation value 
$\big\lag \hat T^\m_{\ \n}\big\rag$ in QFT not only generally violates both classical energy conditions, but also exhibits 
correlation and coherence effects quite different from that of classical point particles.

The presumed existence of a classical BH horizon and trapped surface leads to a number of difficulties in quantum theory. 
Hawking radiation emerging from the BH at temperature $T\!_H \!=\! \hbar c^3/8\p k_B GM$ in a mixed thermal state, together 
with a trapped surface from which no classical matter, radiation or information can escape, implies a breakdown of unitary evolution, 
and a severe BH  `information paradox'~\cite{Hawk:1975, HawkUnit:1976,Preskill:1992,GiddNelson:1992,Page:1993,Mathur:2009,AMPS:2013,Giddings:2013,MottVauPT,Marolf:2017,UnruhWald:2017,AHMST:2020, EM_RegBHs:2023}. The associated BH entropy proportional 
to the surface area of the horizon $A_H$~\cite{Beken:1973}
\vspace{-3mm}
\be
S\!_{BH} = k_B\,\frac{A_H}{4L_{Pl}^2\!} =  \p k_B \left(\frac{\rM}{L_{\rm Pl}}\right)^{\!2} 
 \simeq 10^{77}\, k_B \left(\frac{\! M}{M_{\odot}}\right)^{\!2}
\label{SBH}
\vspace{-2mm}
\ee
involves the square of the ratio of the macroscopic classical Schwarzschild radius $\rM\!=\!2GM/c^2 \simeq \! 3\, (M/M_\odot)$\,km to 
the extremely microscopic quantum Planck length $L_{\rm Pl}\!= \!\!\sqrt{\hbar G/c^3}\! \simeq\!1.62 \times\! 10^{-33}$ cm. 
At some $19$ orders of magnitude larger than the entropy of a stellar progenitor of the mass of the sun $M_{\odot}$, $S\!_{BH}$ 
is not only enormous, but also in conflict with entropy estimates from statistical physics for relativistic stars~\cite{ZelNovbook}. Since Boltzmann's 
formula $S \!=\! k_B \ln W(E)$ relates the entropy of a system logarithmically to its total number of microstates $W(E)$ at fixed 
energy $E$, (\ref{SBH}) implies that of order $\exp(10^{19})$ additional microstates are somehow associated with the horizon.
This is difficult to reconcile with the classical view of a BH horizon as a featureless mathematical boundary only, 
where $T^\m_{\ \n}= 0$, supposedly possessing no local dynamics or degrees of freedom of its own. 

In addition, although the Hawking temperature $T\!_H$ of the radiation far from the BH is very low, the local temperature 
$T\!_{\rm loc}(r)$, extrapolated back to the horizon $r\!\to\! \rM$ by the Tolman relation~(\ref{Tol}) \cite{Tolman:1930},
\begin{wrapfigure} {r}{.4\textwidth}
\vspace{-4mm}
\be
T\!_{\rm loc}(r) =\ \,\frac{T\!_H}{\hspace{-3mm}\sqrt{\raisebox{-1pt}{$1\ $--}\ \frac{\raisebox{-1ex}{$\rM$\!}}{\displaystyle r}}}\hspace{1cm}
\label{Tol}
\vspace{-6mm}
\ee
\end{wrapfigure}
is arbitrarily high. Large blueshifting of local frequencies and energies $\hbar \w_{\rm loc} \!=\! k_B T\!_{\rm loc}$ as 
the null horizon is approached, even becoming transplanckian there, might be expected to have significant backreaction effects 
on the near-horizion spacetime geometry. Yet the assumption of a rigidly fixed classical BH background geometry, neglecting both $
\big\lag T^\m_{\ \n}\big\rag$ and its quantum fluctuations down to $r\!=\!\rM$, is assumed in Hawking's original treatment of 
BH radiance~\cite{Hawk:1975}.

It is important to recognize that the unbounded growth of $T\!_{\rm loc}(r)$ in (\ref{Tol}) and the potentially large quantum effects 
near the BH horizon it suggests, holds for BH's of any size, and does not depend in any way upon large local curvatures at $r=\rM$. Although a 
special (Unruh) state exists for which the renormalized $\big\lag \hat T^\m_{\ \n}\big\rag$ remains small on the future horizon of a
BH~\cite{Unruh:1976}, in many, indeed most, other states $\big\lag T^\m_{\ \n}\big\rag$ is arbitrarily large on the horizon for BH's 
of any mass~\cite{ChrFul:1977,EMVau:2006}. The unbounded growth of $\big\lag\hat T^\m_{\ \n}\big\rag$ as $r\! \to\! \rM$ in such 
states can be traced not to the classical curvature becoming large there, but to the timelike Killing vector $K=\pa/\pa t$ becoming null 
at the horizon, a coordinate invariant (though non-local) property of a BH horizon, independent of local curvature. As $K_\m K^\m\! \to\! 0$
the unambiguous separation of positive and negative energy wave excitations in QFT breaks down, there is no unique `vacuum' 
state, and a large $\big\lag \hat T^\m_{\ \n}\big\rag$ leading to a quantum vacuum phase transition cannot be excluded. 

That $\big\lag \hat T^\m_{\ \n}\big\rag$ grows large on the null horizon in generic states is a consequence of the conformal 
behavior of QFT in the near-horizon limit, and the stress tensor of the conformal anomaly~\cite{EMVau:2006,AndEMVau:2007,EMZak:2010}. 
These previous studies show that the effective action of the conformal anomaly is relevant at macroscopic distance scales, and 
particularly on BH horizons~\cite{EMEFT:2022}. A large $\big\lag \hat T^\m_{\ \n}\big\rag$ in the near-horizon region from QFT 
anomaly effects would result in significant departure of the geometry from that of a classical BH, and the possibility that no trapped 
surface at all is ever formed.
 
In Refs.~\cite{gravastar:2001,MazEMPNAS:2004,Grav_Univ:2023} a non-singular time independent endpoint of gravitational collapse, 
free of all BH puzzles and paradoxes was proposed. This gravitational vacuum condensate star (`gravastar') is a cold, maximally
compact object with an interior non-singular de Sitter $p_{_V}=\!-\r_{_V}$ condensate phase and an exterior Schwarzschild geometry (in 
the non-rotating case), of arbitrary total mass $M$. These two phases are separated by a phase boundary with a positive physical 
surface tension at $r=\rM$where the de Sitter and BH horizon would be otherwise. The de Sitter interior has no entropy, befitting a
macroscopic quantum state at zero temperature, and there is no large BH entropy (\ref{SBH}) to be explained.

Remarkably, just such a solution of the classical Einstein eqs.~is obtained from Schwarzschild's 1916 constant density interior
solution~\cite{Schwarzs:1916}, if taken to its compact limit with radius $r_{\rm star}\!\to\! \rM$~\cite{MazEM:2015}. 
In this limit the Schwarzschild star with a constant $p_{_V}\!=\! -\r_{_V}< 0$ interior is the prototype of a ultra-compact regular `BH' solution.
With a non-singular interior and no trapped surface, it is a concrete counterexample evading the classical singularity theorems, 
with a physical surface at $r\!=\!\rM$ precluding any analytic continuation through a classical horizon. This classical solution 
with its phase boundary at $r=\rM$ instead of a BH horizon demonstrates that a surface and non-singular `BH' interior does 
not require any violation of the Equivalence Principle, as nothing prevents Einstein's elevator from coming to an abrupt end of its 
free fall either on earth or at a surface at $r=\rM$, if a surface exists there.

The classical Schwarzschild star reviewed in Sec.~\ref{Sec:gravcl} has a positive surface tension at $r\!=\!\rM$~\cite{MazEM:2015}, 
but does not provide a microscopic origin for this surface tension. For that, a consistent EFT taking the stress tensor of quantum matter and
the quantum vacuum phase boundary into account is necessary. 

The crucial feature of the conformal anomaly of quantum matter is that it implies the existence of light cone singularities, which may 
be seen as $1/k^2$ massless poles in momentum space correlation functions, even in flat space~\cite{GiaEM:2009,ArmCorRose:2009,TTTCFT:2019}.
These light cone correlations, absent in the classical theory, are relevant at all scales and particularly on null horizons, and are incorporated 
in the effective action and stress tensor of the conformal anomaly, reviewed in Sec.~\ref{Sec:Anom}. The scaling behavior of this effective 
action indicates independently that it is a necessary and relevant addition to the classical Einstein-Hilbert action at macroscopic
scales~\cite{MazEMWeyl:2001,EMSGW:2017}. 

The second, related essential element of the low energy EFT of gravity is a satisfactory formulation of vacuum energy and 
the gravitational vacuum.  The cosmological constant $\La$ has long been associated with vacuum energy, suggesting
its relation to quantum zero point energy~\cite{WeinbergRMP:1989}, while its value implied by the accelerating expansion of the 
universe~\cite{RiessSN:1998,PerlmutterSN:1999} is said to be `unnaturally' small. This conundrum is removed if the rigidly 
fixed $\La$ of classical general relativity is replaced by $\La_{\rm eff}  \!\propto\! F^{\a\b\g\l}F_{\a\b\g\l}$ in terms of an abelian 
$4$-form gauge field strength $F$. Under this replacement there is no fine tuning problem, since the value of $F$ and gravitational
effects of vacuum energy depend upon macroscopic boundary conditions, not zero point energies of QFT in flat space at extremely short distances~\cite{EMEFT:2022}. 

The importance of this replacement for BH's is that although the field $F$ is a constant in the absence of sources in empty flat 
space, (Sec.~\ref{Sec:F}), it can change if there are sources of the `Maxwell' equation~$d \ast F = \ast J$. Just such a $3-$current source for $F$ 
arises in the near-horizon region if $F =dA$ is identified with the exact $4$-form field of the Euler class, with $A$ the 
Chern-Simons $3$-form, defined in terms of the spin connection $\w^a_{\ b}$, and there are massless fermion fields 
minimally coupled to $\w^a_{\ b}$~\cite{Shap:2002}. 

\begin{wrapfigure}{r}{.4\textwidth}
\vspace{-4mm}
\hspace{1cm}
\includegraphics[height=9cm,width=5cm,trim=0cm 0cm 0cm 0cm, clip]{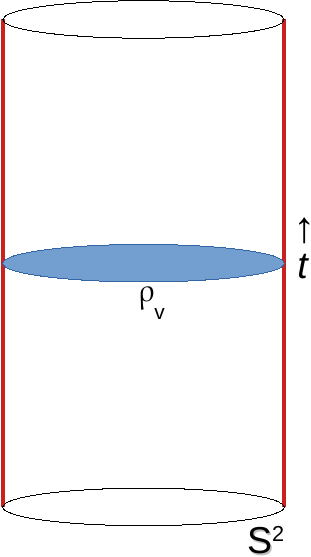}
\vspace{-3mm}
\singlespace
\caption{Worldtube with topology $\mathbb{R}\otimes\mathbb{S}^2$ of the thin boundary layer at $r \!\simeq\! \rM$
where the $3$-current $J$ of (\ref{Jphi}) is large, and the vacuum energy changes rapidly from $\r_{_V}\!>\! 0$ in the 
de Sitter interior (blue shaded) region to $\r_{_V}\!=\!0$ Schwarzschild exterior of a gravastar.}
\label{Fig:Tube}
\vspace{-4mm}
\end{wrapfigure}

Treating $\w^a_{\ b}$ and hence $A$ as independent of the metric is equivalent to admitting torsion in the general description 
of geometry in Einstein-Cartan theory~\cite{Cartan:1922,Hehl:1976}. Since local frequencies are blueshifted near the horizon according 
to (\ref{Tol}), when $\hbar \w\!_{\rm loc}\!> \! m_\n c^2$, exceeding the mass of the lightest fermions, such as neutrinos, these 
fermions can be treated as massless in the near-horizon region and hence contribute to the conformal anomaly. This generates 
a $3$-current $J^{\a\b\g}$ on the $3$-dimensional worldtube of the $r\!\simeq\! \rM$ surface with topology $\mathbb{R}\times \mathbb{S}^2$, 
where $F$ and hence $\La_{\rm eff}$ change rapidly in a boundary layer separating the $\La_{\rm eff} > 0$ interior from 
the $\La_{\rm eff}=0$ exterior: {\it cf.} Fig.~\ref{Fig:Tube}. The activation of torsion and the topological susceptibility that results 
in this region is described in detail in Secs.~\ref{Sec:Tors}-\ref{Sec:Suscep}.

With these essential quantum elements the EFT of low energy gravity provides a first principles basis for gravitational condensate 
stars~\cite{Grav_Univ:2023}, realizing the quantum phase transition hypothesis of \cite{gravastar:2001,MazEMPNAS:2004,ChaplineHLS:2001}, 
in a mean field description. The effective action in the near-horizon region incorporating these elements is given in Sec.~\ref{Sec:EffAct}. 
After a rescaling of variables in Sec.~\ref{Sec:Rescale} appropriate for a thin boundary layer, a variational ansatz and boundary conditions 
for the fields are given in Sec.~\ref{Sec:BoundCond}, indicating that a solution for the phase boundary layer of a non-rotating gravitational 
vacuum condensate star exists in the EFT, the classical limit and prototype of which was inherent already in the 1916 constant density 
Schwarzschild star at its maximum compactness limit $r_{\rm star}\!\to\! \rM$~\cite{MazEM:2015}. 

A summary of the results of the paper is given in Sec.~\ref{Sec:Summary}. There are three appendices collecting 
various detailed derivations and useful formulae referred to in the main text. 

\section{The Schwarzschild Star: Classical Limit of a Gravastar}
\label{Sec:gravcl}

The general static, spherically symmetric geometry is expressed by the line element  
\vspace{-3.5mm}
\be
ds^2 = -f(r)\, dt^2 + \frac{dr^2}{h(r)\!} + r^2\, \big(d\th^2 + \sin^2\!\th \, d\f^2\big)
\label{sphsta}
\vspace{-3.5mm}
\ee
in terms of two functions of the radius $r$, namely $f(r)$ and $h(r)$. The static Killing field $K\!=\!\pa/\pa t$ has the coordinate 
invariant scalar norm $-K^\m K_\m =f(r)$ in the coordinates (\ref{sphsta}) adapted to the symmetries. 

As shown in \cite{MazEM:2015}, the constant density interior solution originally found by Schwarzschild~\cite{Schwarzs:1916}, but 
with its radius $r_{\rm star}$ taken to its compact limit $r_{\rm star}\!\to\! \rM$ is of the form (\ref{sphsta}) with the piecewise continuous 
functions $f$ and $h$ given by
\vspace{-3mm}
\bes
\begin{align}
  f(r) &=
  \begin{cases}
    \displaystyle \dfrac{1}{4}\, \bigg(1- \sdfrac{r^2}{\rM^2}\bigg)\,,
    &  0\le r \le \rM
    \\[1ex]
    \displaystyle 1- \sdfrac{\rM}{r}\,,
    & r \ge \rM
  \end{cases}
  \\
  h(r) &=
  \begin{cases}
    \displaystyle 1- \sdfrac{r^2}{\rM^2}\,,
    & 0\le r \le \rM
    \\[1ex]
    \displaystyle 1- \sdfrac{\rM}{r}
    & r \ge \rM
  \end{cases}\qquad \quad {\rm with}  \qquad \rM = \frac{2GM}{c^2}\,.
\end{align}\\[-7mm]
\label{fhgrav}
\vspace{-2mm}\ees
The line element (\ref{sphsta}) with (\ref{fhgrav}) thus describes an interior static patch of de Sitter space with  
\vspace{-1mm}
\be
\r_{_V} = - p_{_V} = \frac{3}{8\p G\rM^2}\,, \qquad {\rm for} \qquad r< \rM 
\vspace{-2mm}
\label{rhop}
\vspace{-1mm}
\ee
corresponding to a cosmological term of $\La_{\rm eff}=\!3H^2\!=\!3/\rM^2$, joined to a Schwarzschild exterior with 
\vspace{-4mm}
\be
 \r = p =0 \qquad {\rm for} \qquad r > \rM
\label{rhop0}
\vspace{-4mm}
\ee
at their mutual null horizon boundaries at $r\!=\!\rM\!=\!\rH \!=\!1/H$ where $f\!=\!h\!=\!0$. Since $f(r) =-K^\m K_\m$, the radius $r=\rM$
where $f(r) = 0$ is a coordinate invariant singular surface of the spherically symmetric, static geometry. Unlike a BH, the de Sitter interior is 
not an analytic continuation from the exterior.

The ratio of piecewise continuous functions  
\vspace{-4mm}
\begin{align}
  \sqrt{\frac{f}{h} } &=
  \begin{cases}
    1\,,&  0\le r < \rM \\
  \displaystyle  \sdfrac{1}{2}\,,&  r> \rM
  \end{cases}
\label{jdef}
\end{align}\\[-6mm]
is discontinuous at the horizon surface, and the surface gravity
\vspace{-4.5mm}
\begin{align}
\ka_H(r) = \frac{1}{2} \sqrfr{h}{f}\, \frac{df}{dr} &= -\frac{r}{2\rM^2}\, \Theta(\rM -r) + \frac{\rM}{2 r^2}\, \Theta(r-\rM) \to 
  \begin{cases}
      \displaystyle \ka_{H-} = -\frac{1}{2\rM}\,,&  r \to \rM^- \\[2ex]
      \displaystyle \ka_{H+} = + \frac{1}{2\rM}\,,&  r\to \rM^+
  \end{cases}
\label{surfgrav}
\end{align}\\[-1cm]
tends to equal and oposite values as $r\to \rM^{\mp}$ from below or above respectively. 

\begin{figure}[h]
\vspace{-2mm}
\includegraphics[height=5cm,width=8.5cm, trim=0cm 0cm 0cm 0cm, clip]{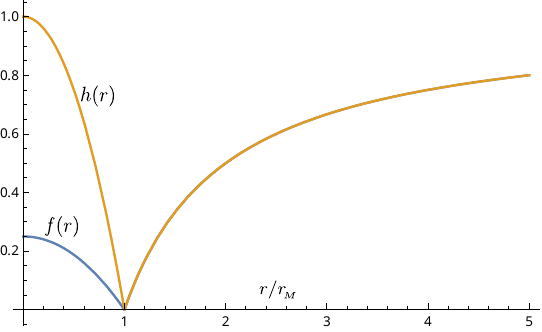}
\hspace{-1cm}
\singlespace
\caption{The metric functions $f(x)$ and $h(x)$ of (\ref{sphsta})-(\ref{fhgrav}) for the classical Schwarzschild star, prototype 
of a non-rotating gravastar, showing the discontinuity of their first derivatives at $r=\rM$.}
\label{Fig:SchwStar}
\vspace{-3mm}
\end{figure}

The discontinuity in $\ka_H$ arising from the combined discontinuity in (\ref{jdef}), and that in the first derivative of $f$ illustrated 
in Fig.~\ref{Fig:SchwStar}, gives rise to a Dirac $\d$-function 
\vspace{-2mm}
\be
\frac{d \ka_{H}}{\!dr} = -\sqrt{\sdfrac{f}{h}}\, R^{tr}_{\ \ tr} 
=[\ka_H] \, \d(r-\rM)  -\frac{1}{2\rM^2}\, \Theta(\rM -r) -  \frac{\rM}{r^3}\, \Theta(r-\rM)
\label{dkapH}
\vspace{-2mm}
\ee
in the $R^{tr}_{\ \ tr}$ curvature component, {\it cf.}~(\ref{Rtrtr}), since the derivative of the Heaviside step function, $\Theta$ 
is the Dirac $\d$-function distribution, where $[\ka_H]\!=\! \ka_{H+} \!-\! \ka_{H-} \!= \!1/\rM$ is the discontinuous change in $\ka_H$.

From the classical Einstein eqs., 
\vspace{-3mm}
\be
\begin{aligned}
G^t_{\ t} & = -8\p G \r\\
G^r_{\ r} & = 8 \p G p\\
G^\th_{\ \th} & = G^\f_{\ \f} = 8 \p G p_\perp
\end{aligned}
\vspace{-3mm}
\ee
one obtains, {\it cf.} (\ref{Einstensor})
\be
\frac{d}{dr} \left(\frac{h}{f}\right) =  \frac{r}{f}\,  \Big( G^t_{\ t} - G^r_{\ r} \Big) = -8 \p G \,\frac{r}{f} \,  \big(\r + p\big)
\label{fhratio}
\ee
showing that the ratio $f/h$ is strictly constant if and only if $\r + p \!=\!0$ and $f\!\neq\!0$, as it is in either de Sitter 
or Schwarzschild geometry, but both conditions fail for (\ref{fhgrav}) at $r\!=\!\rM$, where $f\!=\!h\!=\!0$ and the ratio $f/h$ changes 
discontinuously. In that case the curvature component $R^{tr}_{\ \ \,tr}$ (and only this component) becomes singular, and a 
Dirac $\d$-function (\ref{dkapH}) appears in $\sqrt{f/h}\, R^{tr}_{\ \ \,tr}$ at $r=\rM$.

From (\ref{Einstensor}), this same curvature component appears also in the Einstein tensor densities 
$\sqrt{\frac{f}{h}}\, G^\th_{\ \th} =  \sqrt{\frac{f}{h}}\,  G^\f_{\ \f}$, and therefore corresponds to a surface stress tensor 
$^{(\S )}T^A_{\ \,B}$ 
\vspace{-3mm}
\be
^{(\S )}T^A_{\ \,B}\, \sqrt{\frac{f}{h}} = \cS^A_{\ \,B} \, \d(r-\rM)\,,\qquad  \cS^A_{\ \,B}= \frac{[\ka_H]}{8\p G} \, \d^A_{\ \, B} 
\label{surfTsph}
\vspace{-2mm}
\ee
for the angular components $A, B = \th, \f$~\cite{MazEM:2015}. The surface stress tensor $^{(\S )}T^A_{\ \, B}$ defined by (\ref{surfTsph}) 
is a well-defined distribution when integrated with the $3$-volume invariant measure $d^3\S_t$ at constant $t$, 
\vspace{-3mm}
\be
d^3\S_t = dr\,d\th\, d\f \sqrt{-g} = dr\,d\th\, d\f\sqrt{\frac{f}{h}}\ r^2\sin\th
\vspace{-3mm}
\ee
in the coordinates of (\ref{sphsta}). The stress tensor (\ref{surfTsph}) with the proper volume integration measure 
thus follows directly from the Einstein eqs.~for the piecewise continuous metric (\ref{fhgrav}) \cite{BelGonEM1:2022}. 
The $\d$-function singularity of (\ref{dkapH}) is a perfectly integrable and rigorously defined distributional source when
integrated $\int dr$, which can then be recognized as the positive physical surface tension \cite{MazEM:2015}
\vspace{-3mm}
\be
\t_s =  \frac{[\ka_H]}{8\pi G} = \frac{c^4}{8\p G \rM} =  \frac{Mc^2}{A_H} = \frac{c^6}{16\p G^2 M} > 0
\label{tauS}
\vspace{-2mm}
\ee
of an infinitesimally thin boundary layer at the Schwarzschild radius $r=\rM = 2GM/c^2$. 

This compact limit of Schwarzschild's 1916 constant density interior solution~\cite{Schwarzs:1916}, a Schwarzschild `star,' may be
viewed as the universal classical limit of the gravitational condensate star model proposed in \cite{gravastar:2001,MazEMPNAS:2004}, 
in the sense that when the thickness of the intervening surface layer localized on the horizon is taken to zero, the resulting solution is
independent of any assumptions about the intervening layer equation of state. The surface stress tensor (\ref{surfTsph}) and surface
tension (\ref{tauS}) are completely determined by the matching of the interior and exterior spacetimes on their respective horizons, 
after appropriate generalization of the Lanczos-Israel junction conditions to a null hypersurface~\cite{BelGonEM1:2022}. 

It is noteworthy that the surface gravities (\ref{surfgrav}) so determined are equal and opposite in sign, which is the condition 
of the balancing of the inward and outward gravitational accelerations and forces applied to the surface, and that the surface tension 
$\t_s$ is positive, necessary for the stability of the surface membrane to perturbations increasing the surface area since $dE_s = \t_s \,dA_H> 0$.

Just as a Schwarzschild BH can have any mass, the ultracompact Schwarzschild star (\ref{fhgrav}) is defined for any $M$, 
provided that the de Sitter interior dark energy density is adjustable to $M$ by (\ref{rhop}). Like the arbitrarily thin 
shell gravastar model of \cite{gravastar:2001,MazEMPNAS:2004,Grav_Univ:2023}, this compact limit of a Schwazschild star is a cold, 
zero entropy solution, as determined by the Gibbs relation $p + \r \!= \! sT + \m N \!=  \!0$,  with no conserved charge
(hence no chemical potential: $\m=0$). There is thus no enormous entropy of (\ref{SBH}) to be accounted for in such a compact object, 
and no conflict with statistical thermodynamics. 

Since $f(r) \ge 0$ in (\ref{fhgrav}) everywhere, only touching zero with a cusp at $r=\rM$ where its derivative is discontinuous, 
Fig.~\ref{Fig:SchwStar}, the interior is not an analytic continuation from the exterior, and there is no true trapped surface, so that 
the condition needed for the Penrose singularity theorem \cite{Penrose:1965} is also not met, and the de Sitter interior is non-singular. 
As a result of the existence of a static Killing time $t$ throughout, there is no conflict with unitary quantum evolution and no information 
paradox~\cite{EM_RegBHs:2023}. 

The significant question unanswered to this point is what underlying physics may be responsible for the surface stress 
tensor (\ref{surfTsph}), allowing the vacuum energy $\La_{\rm eff}$ to change abruptly at $r=\rM$ in order to adjust to 
the value (\ref{rhop}) determined by the mass $M$. The discontinuity in $\ka_H$ and resulting $\d$-function in (\ref{dkapH})
is also certainly an idealization of the classical theory, similar to the discontinuity of electric field across an infinitely
thin conducting surface. Just as in that case, one would expect the infinitely thin surface of the classical 
Schwarzschild star to be replaced by a finite thickness boundary layer depending upon a more adequate mesoscopic
description of quantum matter in the layer.

An abrupt change of vacuum energy at $r\!=\!\rM$ requires a quantum phase transition at the horizon scale, independently 
of $M$ and the smallness of local curvature there. That just such a phase transition and change of vacuum energy density at 
$r\!\simeq\! \rM$ arises from the conformal anomaly of quantum fermionic matter coupling to the $4$-form field $F$
that determines vacuum energy through $\La_{\rm eff} \propto F^2$, as proposed in \cite{EMEFT:2022}, will be shown
in the following sections of this paper.

\section{Macroscopic Effects of the Quantum Conformal Anomaly}
\label{Sec:Anom}

EFT methods rely upon the decoupling of heavy degrees of freedom from low energy physics~\cite{AppCar:1975,Manohar:2017,Don:1994,Bur:2004}. 
However quantum anomalies due to the effects of light or massless fields, do not decouple and affect the symmetries of a QFT through 
anomalous Ward Identities at all energy scales. This is expressed by the principle of anomaly matching from UV to low energy 
EFT~\cite{tHooft:1979}. In QCD the axial anomaly and anomalous chiral Ward Identity require a specific Wess-Zumino (WZ) addition 
to the low energy EFT of mesons~\cite{WZ:1971,Leut:1994,WeinbergQFT}, which is not decoupled or suppressed by any high energy scale, 
but on the contrary is responsible for the measured low energy pion decay rate $\p^0\! \to \!2\g$~\cite{TreiJackGross:2015,Bertlbook}. 

One way to see that anomalous Ward identities can have significant effects at energies far below the Planck scale is that they imply $1/k^2$ 
massless poles in momentum space correlation functions~\cite{GiaEM:2009,ArmCorRose:2009,BlaCabEM:2014}. The conformal anomaly 
in the trace of the stress-energy tensor and resulting anomalous Ward identity has been shown to be responsible for the $1/k^2$ pole in
 $\big\lag \hat  T^{\a\b} \hat  T^{\g\l}  \hat T^{\m\n}\big\rag$~\cite{TTTCFT:2019}. This $1/k^2$ anomaly pole describe 
correlations on light cones over macroscopic distances, and signals the existence of a massless scalar excitation that is not present 
in classical GR, nor in any expansion in higher order local curvature invariants. Such light cone correlations are particularly relevant 
on null horizons, where the anomaly has also been shown to be responsible for large $\big\lag \hat T^\m_{\ \n}\big\rag$ in generic 
states on both BH and cosmological horizons~\cite{EMVau:2006,AndEMVau:2007,EMZak:2010}, as well as in a 2D model of 
gravitational collapse~\cite{QuanGrav2D:2023}.
 
The conformal anomaly in the presence of background curvature or gauge fields appears first in the non-zero expectation 
value of the trace of the renormalized energy-momentum tensor~\cite{CapperDuff:1974,DeWitt:1975,Duff:1977,BirDav}\pagebreak
\vspace{-5mm}
\be
\big\lag \hat T^\m_{\ \,\m} \big\rag\ = a\,\Big( E - \tfrac{2}{3} \sq R\Big)  +  b\, C^2 + \sumi \ \b_i\, \cL_i
 \equiv \ \frac{\cA}{\!\!\!\sqrt{-g}}
\label{tranom}
\vspace{-4mm}
\ee
when the underlying classical theory is scale and conformally invariant, and one might have expected this trace to vanish. 
In (\ref{tranom}), $\cA = \!\!\sqrt{-g}\, \big\lag \hat T^\m_{\ \,\m} \big\rag$ is the anomaly density, and
\vspace{-4mm}
\be
E\,  =R_{\a\b\g\l}R^{\a\b\g\l}-4R_{\a\b}R^{\a\b} + R^2\,,\qquad\quad
C^2\! = C_{\a\b\g\l}C^{\a\b\g\l}\! = R_{\a\b\g\l}R^{\a\b\g\l} -2 R_{\a\b}R^{\a\b}  + \sdfrac{1}{3}R^2
\label{ECdef}
\vspace{-3mm}
\ee
are the Euler class invariant and the square of the Weyl conformal tensor respectively. The $\cL_i$ are dimension-four invariants 
of gauge fields, such as $\cL_F\!=\! F_{\a\b}F^{\a\b}$ for coupling to electromagnetism or $\cL_G\!=\! {\rm tr}\, \{G_{\a\b}G^{\a\b}\}$ 
for coupling to the gluonic gauge fields of QCD, each with coefficients determined by the $\b$-function of the corresponding gauge 
coupling. The $a, b,\b_i$ coefficients in (\ref{tranom}) are dimensionless numbers in units of $\hbar$:
\vspace{-3mm}
\be
a = -\frac{\hbar}{(4 \pi)^2\!}\, \frac{1}{360}\,\Big(N_S + 11 N_F + 62 N_V\Big)\,,
\qquad 
b = \frac{\hbar}{(4 \pi)^2\!}\,\frac{1}{120}\, \Big(N_S + 6 N_F + 12 N_V\Big)\,,
\label{bbprime}
\vspace{-3mm}
\ee
where $(N_S, N_F, N_V)$ are the number of massless conformal scalar, Dirac fermion, and gauge vector fields respectively
\cite{Duff:1977,BirDav}.\footnote{There seems to be no universally accepted standard notation for the 
anomaly coefficients $a$ and $b$. Here the conventions of \cite{EMSGW:2017,EMEFT:2022} are utilized, except $b' \!\to \! a$ in 
recognition of the unique and primary role the topological Euler class plays.} The anomaly coefficients 
(\ref{bbprime}) do not involve the ultrashort Planck scale $L_{Pl}$, and cannot be changed or removed by any local renormalization
counterterm, since they depend upon the number of massless (or effectively massless) fields of each spin, which is a property 
of the low energy matter/radiation content of the SM fields, minimally coupled to curved spacetime by the Equivalence Principle. 
In (\ref{tranom})-(\ref{bbprime}) the spacetime metric itself is treated as a classical field.

As a corollary to the absence of any local counteterm that can remove the $a$ and $b$ terms of the anomaly, it is intrinsically 
a non-local quantum effect, corresponding to a non-trivial cocycle of the local Weyl group~\cite{MazEMWeyl:2001}. An independent 
$\sq R$ term has been omitted from (\ref{tranom}) since it can be altered by a local $R^2$ counterterm in the effective action, 
corresponding to a trivial Weyl cocycle, subject to UV renormalization, and therefore is not a true anomaly. However, the linear combination
\vspace{-4mm}
\be
\sqrt{-g}\,\left(E - \tfrac{2}{3}\sq R\right) \to \sqrt{- g}\, \left(E - \tfrac{2}{3}\sq R\right) + 4\sqrt{-g}\,\D_4\,\s 
\label{Esig}
\vspace{-3mm}
\ee
in $\cA$ transforms simply under the local conformal transformation of the metric $g_{\m\n}\!\to\! e^{2\s}g_{\m\n}$, where
\vspace{-3mm}
\be
\D_4 \equiv \na_\m \left(\na^\m\na^\n +2R^{\m\n} - \tfrac{2}{3} R g^{\m\n} \right)\na_\n
=\sq^2 + 2 R^{\m\n}\na_\m\na_\n - \tfrac{2}{3} R \sq + \tfrac{1}{3} (\na^\m R)\na_\n
\label{Deldef}
\vspace{-3mm}
\ee
is the (unique) fourth order scalar differential operator that is conformally covariant~\cite{Panietz,Rieg:1984},
whose characteristic surfaces are light cones. The other terms in the anomaly density $\cA$ are invariant under a local
conformal transformation of the metric. Thus, grouping a part of the trivial $\sq R$ term with $E$ with the specific
$-2/3$ coefficient in (\ref{Esig}) exhibits the properties of the anomalous terms in (\ref{tranom}) 
under conformal transformations most clearly and simply.

The minimal non-local form of the anomaly effective action is~\cite{Rieg:1984,FradTsey:1984,AntMazEM:1992,MazEMWeyl:2001,BuchOdinShap}
\vspace{-3mm}
\be
S^{NL}_{\! \cA}[g] =\frac {1}{4}\!\int \!d^4\!x\sqrt{-g_x}\, \left(E - \tfrac{2}{3}\sq R\right)_{\!x} \!
\int\! d^4\!y\sqrt{-g_y}\,G_4(x,y)\,\bigg\{\frac{a}{2} \left(E - \tfrac{2}{3}\sq R\right) 
+  b\,C^2 + \sumi\, \b_i\,\cL_i\bigg\}_y 
\label{Snonl}
\vspace{-2mm}
\ee
where $G_4(x,y)$ is the Green's function inverse of the fourth order differential operator (\ref{Deldef}) between the spacetime 
points $x$ and $y$, satisfying
\vspace{-3mm}
\be
\D_4\, G_4 (x,y) = \d_g^4(x,y) \equiv \frac{1}{\!\!\sqrt{-g}}\,  \bar\d^4(x,y)\,,\qquad 
\int\!d^4y \!\sqrt{-g_y}\,  \D_4\, G_4 (x,y)\, \F(y) = \F(x)
\label{D4del}
\vspace{-3mm}
\ee
where $\bar\d^4(x,y)$ is the scalar density defined by the properties $\bar\d^4(x,y) = 0$ for $x\neq y$, and $\int\! d^4y\, \bar\d^4(x,y) =1$,
with (\ref{D4del}) holding for any continuous scalar function $\F(x)$. The Green's function $G_4$ is singular on the
light cone which is the source of its significance for near-horizon physics of BH's. It is defined up to freedom to add homogeneous 
solutions of $\D_4 \vf = 0$ that must fixed by suitable boundary conditions. 

Although a non-local quantum effect, necessitating the non-local effective action (\ref{Snonl}) in terms of solely the original spacetime 
metric variables, the effective action of the anomaly can nevertheless be rendered in the fully invariant local form~\cite{EMSGW:2017,EMEFT:2022}:
\vspace{-2mm}
\be
\begin{aligned}
S_{\!\!\cA} [g;\vf] &=  -\frac{a}{2}\!\int\!\!d^4\!x\! \sqrt{-g}\,\bigg\{\big(\sq \vf\big)^2  
- 2\,\Big(R^{\m\n}\! - \!\tfrac{1}{3} Rg^{\m\n}\Big) \,(\pa_\m \vf)\,(\pa_\n\vf)\bigg\}  
+ \frac{1}{2} \!\int\!\!d^4\!x\,\cA\,\vf\\
&\equiv  S\!_\cA^{(2)}[g;\vf] +  S\!_\cA^{(1)}[g;\vf]
\end{aligned}
\label{Sanom}
\vspace{-2mm}
\ee
by the introduction of a single new local field $\vf$, where $S\!_\cA^{(2)}$ and $S\!_\cA^{(1)}$ denote the terms in $S_{\!\!\cA} [g;\vf] $
quadratic and linear in $\vf$ respectively. This scalar field can be termed a {\it conformalon} field since it relates geometries in the same 
conformal equivalency class by local conformal transformations and the WZ consistency condition
satisfied by $S_{\!\!\cA}$ \cite{WZ:1971,AntMazEM:1992,MazEMWeyl:2001,EMSGW:2017}:
\vspace{-4mm}
\be
S_{\!\!\cA} [e^{-2 \s} g; \vf] = S_{\!\!\cA}[g; \vf + 2 \s] - S_{\!\!\cA}[g; 2 \s]\,.
\label{SanomWZ}
\vspace{-4mm}
\ee
Variation of $S_{\!\!\cA}$ with respect to  $\vf$ yields the linear eq.
\vspace{-3mm}
\be
\D_4\, \vf  = \frac{1}{2a}\, \frac{\cA}{\!\!\!\sqrt{-g}} = \frac{1}{2}E  -\frac{1}{3}\sq R  +  \frac{b}{2 a } \, C^2 
+  \frac{1}{2a}\ \sumi \ \b_i\, \cL_i
\label{Del4}
\vspace{-2mm}
\ee
which when solved for $\vf$ in terms of the Green's function $G_4(x,y)$ of $\D_4$, and substituted back into (\ref{Sanom}) 
returns the non-local form of the effective action \cite{Rieg:1984,FradTsey:1984,AntMazEM:1992,MazEMWeyl:2001,BuchOdinShap}, 
up to a Weyl invariant term. 

The stress-energy tensor
\vspace{-1mm}
\be
T_{\!\!\cA}^{\m\n}\equiv \frac{2\!}{\!\!\!\sqrt{-g}} \, \frac{\d}{\d g_{\m\n}} \ S_{\!\!\cA}[g;\vf]
\label{Tphi}
\ee
following from (\ref{Sanom}) is covariantly conserved and its trace  $g_{\m\n}T_{\!\!\cA}^{\m\n}[g;\vf] 
= \big\lag \hat T^\m_{\ \ \m} \big\rag$ is (\ref{tranom}), so that the non-local effects of the quantum conformal anomaly 
are taken into account by the local scalar $\vf$ treated classically in (\ref{Tphi}). The explicit general 
form of $T_{\!\!\cA}^{\m\n}[g;\vf] $ is given in Appendix \ref{App:TAnom}.

To see the relevance of the anomaly stress tensor (\ref{Tphi}) at BH horizons one can solve the linear eq.~(\ref{Del4}) for 
$\vf = \vf(r)$ in the classical Schwarzschild geometry (with $\b_i = 0$), resulting in~\cite{EMVau:2006}
\vspace{-3mm}
\be
\vf_S(r) = c_{_S} \ln \left(1- \sdfrac{\rM}{r}\right) + c_{\infty} + \left(1 + \sdfrac{\!b}{a }\right) \,\vf_{\rm inh}\left(\sdfrac{\!r}{\rM}\right) 
\label{phiS}
\vspace{-3mm}
\ee
where $c_{_S}$ and $c_{\infty}$ are integration constants of the homogeoneous solutions of (\ref{Del4}), while 
\vspace{-3mm}
\be
\hspace{-3mm} \vf_{\rm inh}(\br) = -\sdfrac{1}{3}
\left\{ \br +  \Big(\br^2 + 2\br -3 -2\ln \br\Big) \ln \left(\!1 - \sdfrac{1}{\br}\right) + 4\, {\rm Li}_2\left(\sdfrac{1}{\br}\right) 
+ 2\, {\rm Li}_2\left(\!1 - \sdfrac{1}{\br}\right) -1  - \sdfrac{2 \p^2}{3}\right\}
\label{vfinh}
\vspace{-3mm}
\ee
when multiplied by $(1 + b/a )$, provides a particular inhomogenous solution of (\ref{Del4}) in the exterior Schwarzschild geometry. 
In (\ref{vfinh}) $\br \equiv r/\rM$ is a dimensionless radial variable, and
\vspace{-2mm}
\be
{\rm Li}_2(z) = -\! \int_0^z \frac{du}{u}\,\ln \big(1-u\big) \ =\  \sum_{n=1}^{\infty} \frac{z^n}{n^2}\,,\quad |z|\le 1
\label{Dilog}
\vspace{-2mm}
\ee
is the dilogarithm (Spence) function. The particular solution (\ref{vfinh}) has been chosen to vanish at $r\!=\!\rM$ and 
be finite as $r\!\to\! \infty$. Two additional solutions of the fourth order homogeneous eq.~(\ref{Del4}), namely
\bes
\vspace{-4mm}
\begin{align}
&\hspace{3cm}\vf_2(r) = \br^2 + 2\br + 2\ln\br  \\[-1.8ex]
&\hspace{-3.3cm}{\rm and}\nonumber \\[-1.5ex]
&\vf_1(r)=  2 \br +6 \ln\br - \left(\br^2 + 2\br -3 - 4\ln\br\right) \ln \left(\!1 - \sdfrac{1}{\br}\right)
- 4\, {\rm Li}_2\left(\sdfrac{1}{\br}\right) 
\end{align}
\vspace{-1.1cm}
\label{phi12}\ees

\noindent 
that diverge as $r^2$ and $r$ respectively as $r\to\infty$ have been excluded from (\ref{phiS}) by the requirement of bounded 
total energy and asymptotic flatness. These solutions are relevant for cosmological or other spacetimes which are not asymptotically flat.

The derivative of (\ref{phiS}) is expressible in terms of elementary functions,
\vspace{-4mm}
\be
\frac{d \vf_S}{\!dr} =\frac{c_{_S} \rM}{r(r-\rM)} - \left(1 + \frac{\!b}{a }\right) 
\left\{\frac{2}{3\rM} \left(\frac{\!r}{\rM} + 1 + \frac{\rM}{r} \right) \ln \left(1-\frac{\rM}{r}\right)   
+ \frac{2}{3\rM} + \frac{1}{r} \right\}
\label{dphiS}
\vspace{-3mm}
\ee
while the asymptotic forms of (\ref{phiS}) are
\vspace{-2mm}
\be
\vf_S(r) \to \left\{ \begin{array}{lr} c_{_S}\ln \left(\sdfrac{\!r}{\,\rM}-1\right) + c_\infty &\\[1ex]
\hspace{5mm} - \left(\sdfrac{\!r}{\,\rM}-1\right) \left\{ c_{_S} + \left(1 + \sdfrac{\!b}{a }\right) 
\left[2\ln   \left(\sdfrac{\!r}{\,\rM}-1\right) - \sdfrac{1}{3}\right] + \dots \right\}, &r \to \rM\\[2ex]
c_\infty + \sdfrac{1}{3} \left(1 + \sdfrac{\!b}{a }\right) \left\{ \sdfrac{7}{2} + \sdfrac{\p^2}{3}
- \sdfrac{11}{3}\sdfrac{\rM}{r}  - \sdfrac{13}{12} \left( \sdfrac{\rM}{r} \right)^2+ \dots\right\}
- c_{_S} \sdfrac{\rM}{r}  \left\{1 +  \sdfrac{\rM}{2r} + \dots\right\}, &r \to \infty
\end{array}\right.
\label{phiSlim}
\vspace{-2mm}
\ee
at the horizon and infinity respectively. The leading divergence at $r=\rM$ depends upon the state dependent 
integration constant $c_{_S}$, which is generically non-zero, while (\ref{dphiS}) shows that $\vf_S'$ contains a subleading logarithmic 
divergence from $\vf_{\rm inh}$ that is independent of $c_{_S}$. This subleading divergence cannot be removed, except at the price of involving 
one or both of the solutions (\ref{phi12}) which diverge as $r \to \infty$, or introducing time dependence 
in $\vf$ \cite{EMVau:2006}, excluded here by restriction to a static solution.

If the general spherically symmetric static solution (\ref{phiS}) for $r > \rM$ in the Schwarzschild background, regular at infinity, 
is substituted into the expression for the anomaly stress tensor (\ref{Tphi}), (\ref{Ttens})-(\ref{CFterms}), one finds~\cite{EMVau:2006,EMZak:2010}
\vspace{-9mm}
  \be
\big(T^\m_{ \ \,\n}\big)_\cA \to \frac{c_{_S}^2}{2\rM^2}\ \frac{a }{(r-\rM)^2}\,
\left(\begin{array}{cccc} -1& 0 & 0& 0\\ 0 &\sdfrac{1}{3} & 0 & 0 \\ 0 & 0 &\sdfrac{1}{3}& 0 \\ 0 & 0 &0 &\sdfrac{1}{3}
\end{array}\right)
 \to \infty
\qquad {\rm as} \qquad r \to \rM
\label{ThorizM}
\vspace{-3.5mm} 
\ee
which is proportional to the scalar invariant  $(-K_\m K^\m)^{-2} = f^{-2}$ and quadratically divergent on the Schwarzshild BH horizon 
for any $c_{_S} \neq 0$, and for a BH of any size and mass $M$.

The behavior (\ref{ThorizM}) and its derivation from the conformal anomaly effective action is a consequence of the extreme blueshifting 
of local frequencies and energies (\ref{Tol}), which renders all finite mass scales irrelevant as $r\! \to\! \rM$~\cite{EMZak:2010}, and
results in conformal scaling behavior of the near horizon geometry~\cite{SacSol:2001}. Since the stress tensor is a dimension four, 
conformal weight four operator, it behaves generically as the fourth power of (\ref{Tol}) in (\ref{ThorizM}), {\it i.e.} 
$\w_{\rm loc}^4 \propto f^{-2}$. Noting that $\vf$ is a scalar, as is the norm of the static Killing field $K\!=\!\pa_t$, and 
$\sqrt{-K^\m K_\m}\!=\!\sqrt{f(r)}$, the divergence of (\ref{ThorizM}) depending on the inverse fourth power of this norm as $f\to 0$ 
on the horizon is also a coordinate invariant scalar, independent of the curvature or the Planck scale. Even if $c_{_S}$ is tuned 
to be zero identically, the subdominant term in (\ref{phiSlim}) leads to $(r-\rM)^{-1}, \ln^2 (r-\rM)$ and $\ln (r-\rM)$ growing 
terms at the horizon. There is no static, spherically symmetric solution $\vf = \vf(r)$ of (\ref{Del4}) with a finite stress tensor
at $r=\rM$ that also falls off faster than $1/r^2$ as $r\to \infty$, and hence possesses finite total energy.

From (\ref{bbprime}) since $a  < 0$ for all SM quantum fields, (\ref{ThorizM}) is a negative local energy 
density and pressure at the horizon, which violates all classical energy conditions. Since (\ref{ThorizM}) becomes arbitrarily 
large and negative as $r\!\to\!\rM$, it signals a vacuum instability, with significant effects on the classical geometry
sufficiently close to the horizon but before the horizon and any actual divergence in $T^\m_{ \ \,\n}$ is reached. 
In fact since the classical terms in Einstein's eqs.~are of order $1/\rM^2$, the anomaly stress tensor (\ref{ThorizM})
$8 \p G\, (T^\m_{\ \n})_\cA \propto L_{\rm Pl}^2/\rM^4 f^2$ becomes comparable to the classical terms when 
\vspace{-4mm} 
\be
f = \left(1 - \sdfrac{\rM}{r}\right) \, \lesssim  \e \equiv \frac{L_{\rm Pl}}{\rM}  
\simeq 5.47 \times 10^{-39}\  \frac{M_\odot}{\!M}\ll 1
\label{epsdef}
\vspace{-4mm}
\ee
is satisfied. This extreme near-horizon region defined by $r\! \in\! (\rM \!- \!L_{\rm Pl}, \rM\! +\! L_{\rm Pl})$ of radial coordinate width 
$\D r \!\simeq\! L_{\rm Pl}$ of the Planck length has a physical thickness in the Schwarzschild metric of
\vspace{-3mm}
\be
\ell = \int_{\rM}^{\rM + L_{\rm Pl}} \!\! dr\, \left(1 - \sdfrac{\rM}{r}\right)^{-\frac{1}{2}}\simeq  2\! \sqrt{\rM L_{\rm Pl}} \ 
\simeq \ 4.37\,\times\,10^{-14}\!\sqrt{\!\frac{\! M}{M_\odot}\!} \ \, {\rm cm}\ \gg\ L_{\rm Pl}\,.
\label{elldef}
\vspace{-2mm}
\ee
Thus although $\ell \ll \rM$ is very small on the macroscopic astrophysical scale of $\rM$, it is still very much greater than 
the Planck scale $L_{\rm Pl}$, where a fully quantum treatment of gravity would be needed. The low energy EFT of gravity 
and a semi-classical mean field treatment of the gravitational metric field, taking the stress tensor of the conformal anomaly 
into account, should remain valid within the boundary layer of physical width $\ell \gg L_{\rm Pl}$ near the horizon.

Similar and analogous considerations apply to the static de Sitter patch interior (\ref{fhgrav}) of a gravastar. In this case the general
solution of (\ref{Del4}) for $\vf = \vf (r)$ is
\vspace{-2mm}
\be
\vf_{dS}(r) = 2 \ln \left(\frac{1+ Hr}{2}\right)+ c_0 + c_q \left(1 - \frac{1}{Hr}\right)  \ln  \left(\frac{1-Hr }{1 + Hr}\right)
+ \frac{c_{_H}}{Hr}\ln  \left(\frac{1-H r }{1 + Hr}\right)\,, \quad r < \rH = \sdfrac{1}{H}
\label{phidS}
\vspace{-2mm}
\ee
where the first term is a particular solution of the inhomogeneous eq.~(\ref{Del4}) chosen to be finite and regular at 
both $r=0$ and $r=\rM$, and the three constants of integration $(c_0,c_q,c_{_H})$ multiply three linearly independent solutions of the
corresponding homogeneous equation that are finite and regular at the origin.  The fourth linearly independent solution of the 
homogeneous differential eq.~(\ref{Del4}), {\it i.e.} $\vf_{-1}(r) = 1/Hr$ has been excluded from (\ref{phidS}) because it is singular 
and leads to a singular stress tensor at $r=0$. The last term in (\ref{phidS}) with $c_{_H} \!\neq \!0$ is the leading divergence on the 
de Sitter static horizon $r \!=\! 1/H$, again proportional to $\ln f(r)$ for $f(r) \to 0$  

The solution (\ref{phidS}) has the limits
\vspace{-3mm}
\bes
\begin{align}
\vf_{dS}(r) &\to \left\{ \begin{array}{ll} c_0 -2\,\big( c_q + c_{_H} + \ln 2\big) + 2\, \big(1 -c_q\big)\, Hr + \dots \ , &r \to 0 \\[1ex]
{\displaystyle c_{_H} \ln \left(\sdfrac{1-Hr}{2}\right)  + c_0 +\big(c_{_H} - c_q\big) \big(1-Hr\big)\ln \left(\sdfrac{1-Hr}{2}\right) } +  \dots\ , 
&r \to \rH =\displaystyle \sdfrac{1}{H}
\end{array}\right.\\
\frac{1}{H} \frac{d \vf_{dS}}{\!dr}&\to - \frac{c_{_H}}{1-Hr} +\big (c_q-c_{_H}\big)\ln\, (1-Hr) 
+ \dots \ , \hspace{1.3cm} r \to \rH =\displaystyle \sdfrac{1}{H}
\end{align}
\label{phidSlim}
\ees
\vspace{-1cm}

\noindent
which is regular at the origin, but with a logarithmic behavior at the horizon, analogous to (\ref{phiSlim}).
Substituting (\ref{phidS}) in the anomaly stress tensor (\ref{Tphi}), one finds~\cite{EMVau:2006,EMZak:2010}
\vspace{-3.5mm}
 \be
\big(T^\m_{ \ \,\n}\big)_\cA \to 2\, c_{_H}^2\, H^4 \ \frac{a }{(1 -Hr)^2}\,
\left(\begin{array}{cccc} -1& 0 & 0& 0\\ 0 &\sdfrac{1}{3} & 0 & 0 \\ 0 & 0 &\sdfrac{1}{3}& 0 \\ 0 & 0 &0 &\sdfrac{1}{3}
\end{array}\right)
 \to \infty
\qquad {\rm as} \qquad r \to \rH = \sdfrac{1}{H}
\label{ThorizH}
\vspace{-2.5mm}
\ee
which also diverges quadratically on the de Sitter static horizon. If one makes use of (\ref{fhgrav}) and (\ref{surfgrav}),
and notes that in each case (\ref{phiS}) and ({\ref{phidS}) the divergent solution for $\vf$ behaves as $c_{_S} \ln f$ or  $c_{_H}\ln f$
on their respective horizons,  the factor multiplying the stress tensor matrices in (\ref{ThorizM}) and (\ref{ThorizH}) is 
$4\, c^2_{\pm} \ka_{H\pm}^4 a /f^2$, where the plus $(+)$ sign refers to the exterior Schwarzschild values of $c_{_S},$ and
surface gravity $\ka_{H+}= 1/2\rM$ while the minus $(-)$ sign refers to the corresponding interior de Sitter values 
of $c_{_H}, \ka_{H-}= H/2 = 1/2\rM$ of the classical gravastar, joined at their mutual horizons $\rH= H^{-1} = \rM$.

Both (\ref{ThorizM}) and (\ref{ThorizH}) illustrate that the stress-energy tensor of the conformal anomaly can become arbitrarily 
large in the vicinity of a BH or de Sitter static event horizon, even dominating the classical terms, irrespective of how small the local
curvature is there. On general grounds of its scaling behavior, the WZ action (\ref{Sanom}) of the conformal anomaly is a relevant 
addition to the classical Einstein-Hilbert action for the low energy EFT of gravity, just as the WZ action of the axial anomaly is relevant 
to the low energy EFT of mesons~\cite{MazEMWeyl:2001,EMSGW:2017,EMEFT:2022}. Since these non-local effects are the result of 
light cone singularities associated with the anomaly, they become particularly pronounced on event horizons where QFT becomes
conformal, and the static Killing vector $\pa/\pa t$ becomes null. 

\section{The $\bf 4$-Form Vacuum Energy Condensate} 
\label{Sec:F}

The second necessary element of the low energy EFT of gravity is a consistent characterization of vacuum energy, independent
of ultrashort distance physics. It has been noted~\cite{Aurilia:1978,DuffvanN:1980,AurNicTown:1980,HenTeit:1984, HenTeit:1986}
that a positive cosmological term of classical GR is equivalent to a $4$-form abelian gauge field strength
\vspace{-4mm} 
\be
F = \sdfrac{1}{\,4!}\, F\!_{\a\b\g\l}\, dx^\a \wedge dx^\b \wedge dx^\g \wedge dx^\l 
\label{F4}
\vspace{-4mm}
\ee
provided that $F$ is exact, {\it i.e.} the curl of a totally anti-symmetric $3$-form gauge potential, 
\vspace{-4mm}
\begin{align}
&\hspace{2cm}F=dA\,,\qquad\qquad A = \sdfrac{1}{\,3!} \, A_{\a\b\g} \ dx^\a \wedge dx^\b \wedge dx^\g\nn
&F\!_{\a\b\g\l} =4\, \pa_{[\a} A_{\b\g\l]} = 4\, \na_{[\a} A_{\b\g\l]} = \na_{\a} A_{\b\g\l} - \na_{\b} A_{\a\g\l}+ 
 \na_{\g}A_{\a\b\l}- \na_{\l} A_{\a\b\g}\,,
\label{Fexact}
\end{align}
\vspace{-1.2cm}

\noindent
and provided that $F$ is supplied with the `Maxwell' action
\vspace{-3mm}
\be
S\!_F[g,A] =  -\sdfrac{1}{2\vk^4}\int F \wedge \ast F=  -\sdfrac{1}{2 \times\,4!\, \vk^4}\! \int d^4\!x \sqrt{-g} \  F\!_{\a\b\g\l}F^{\a\b\g\l}
= +\sdfrac{1}{2\, \vk^4}\! \int d^4\!x \sqrt{-g} \,\tF^2
\label{Maxw}
\vspace{-2mm}
\ee 
of $3$-form electrodynamics, in generalization of electrodynamics of the $1$-form vector potential $A_\m$.

In (\ref{Maxw}) $\ast F \equiv \tF$ is the Hodge dual of $F$.
\vspace{-3mm}
\be
\ast F \equiv \tF = \frac{1}{\,4!}\, \ve^{\a\b\g\l}\, F_{\a\b\g\l} \qquad {\rm so \ that}\qquad  F^{\a\b\g\l} = -\ve^{\a\b\g\l}\,\tF
\label{Fdual}
\vspace{-3mm}
\ee
where $\ve_{\a\b\g\l}$ is the Levi-Civita tensor in curved space, defined by (\ref{vol4form})-(\ref{Eps}). The negative sign in the 
second relation of (\ref{Fdual}) results from the Lorentzian signature metric $(-+++)$, so that $\ve_{\a\b\g\l}\ve^{\a\b\g\l} = -24$.

The equivalence of (\ref{F4})-(\ref{Maxw}) to a cosmological term follows from two special properties of a $D$-form $F$
matched to the number of $D\!=\!4$ spacetime dimensions, namely, in the absence of sources,
\begin{enumerate}[label=(\roman*), topsep=0pt,itemsep=-6pt,partopsep=6pt, parsep=4pt]
\item $F$ is constrained to be a constant, with no propagating degrees of freedom, and 
\item its stress tensor is proportional to the metric $g^{\m\n}$, hence equivalent to a vacuum energy.
\end{enumerate}

The simplest example of this is in $D=2$, where the anti-symmetric $2$-form Faraday-Maxwell field strength has 
but one non-vanishing component, $F_{10} \!= \!-F_{01} \!= \!E$, namely the electric field in one spatial dimension. 
Since $E$ satisfies the Gauss Law constraint $\pa_x E =  0$ and the Maxwell eq.~$\pa_t E = 0$ in the absence of any 
sources, it is a non-propagating field, and a spacetime constant. The Maxwell stress tensor in $D=2$ is 
$- g^{\m\n} E^2/2 e^2$,  and thus equivalent to a positive cosmological term $\La > 0$ for any $E \neq 0$. The absolute 
ground state energy is attained if and only if $E=0$ in flat space.

Likewise in $D=4$ the first property (i) follows from the `Maxwell' eq.
\vspace{-3mm}
\be
\na\!_\l F^{\a\b\g\l} \,= 0 \quad {\rm or} \quad \pa_\m \tF = 0 \qquad {\rm for} \qquad J^{\a\b\g}=0
\label{FnoJ}
\vspace{-3mm}
\ee
obtained by varying the free action $S\!_F$ with respect to $A_{\a\b\g}$ in the absence of any sources, while the second 
property (ii) follows from the energy-momentum-stress tensor 
\vspace{-2mm}
\be
T^{\m\n}_{\! F} = \frac{2\!\!}{\!\sqrt{-g}} \frac{\d S\!_F}{\d g_{\m\n}} = 
-\frac{1}{4!\, \vk^4}\left(\sdfrac{1}{2}\, g^{\m\n} F^{\a\b\g\l}F_{\a\b\g\l} - 4 F^{\m\a\b\g}F^{\n}\!_{\a\b\g}\right)
= -\frac{1} {2\vk^4}\, g^{\m\n} \,\tF^{\,2}
\label{TF}
\vspace{-2mm}
\ee
obtained by varying $S\!_F$ with respect to the metric, and making use of the complete anti-symmetry of $F_{\a\b\g\l}$ with respect to 
all of its indices.  Thus $\tF= F^{0123}$ together with any anti-symmetric permutation of indices is the only non-zero component
of $F$, and this `electric' component is a constant in flat space from (\ref{FnoJ}),  analogous to $E\!=\! F^{01}$ in $D\!=\!2$. The parameter
$\vk$ in (\ref{Maxw}) and (\ref{TF}) is a free parameter with dimensions of mass, like the electric coupling $e$ in $D=2$,
whose significance as the topological susceptibility of the gravitational vacuum is analogous to electric charge screening
in the $2D$ Schwinger model~\cite{Seiler:2002}, and discussed in Sec.~\ref{Sec:Tors}.

It follows from (i)-(ii) or (\ref{FnoJ})-(\ref{TF}) that the $4$-form field strength with `Maxwell' action (\ref{Maxw})
is equivalent to a cosmological term in Einstein's eqs.~in $D\!=\!4$, with the identification
\vspace{-3mm}
\be
\La_{\rm eff} = \frac{4\p G}{\vk^4}\, \tF^{\,2} \ge 0
\label{Lameff}
\vspace{-3mm}
\ee
the effective (necessarily non-negative) cosmological constant, for $\vk$ and $\tF$ real constants. 
In this way one can freely trade the constant $\La$ of classical GR for a new fundamental constant $\vk$ 
of the low energy EFT, together with an integration constant of the constraint $\pa_\m \tF= 0$, obtained from (\ref{FnoJ}).

The replacement of the constant $\La$  of classical GR by a constant $\La_{\rm eff}$ of (\ref{Lameff}) although quite simple, has
several significant consequences. First, like the 2D example, only strictly non-negative $\La_{\rm eff} \ge 0$ and de Sitter-like regions
are possible. Second, as a corollary, the absolute minimum of energy $\La_{\rm eff} = 0$ is attained for zero field strength,
if and only if the integration constant $\tF = F^{0123}=0$. Third, this zero value of the integration constant $\tF$ is both allowed 
and {\it required} by consistency with the sourcefree Einstein's equations
\vspace{-3mm}
\be
\left[R_{\m\n} - \sdfrac{R}{2} g_{\m\n}\right]_{\rm flat} = 0 =-\La_{\rm eff}\Big\vert_{\rm flat}\,\h_{\m\n}
\label{Einflat}
\vspace{-1mm}
\ee
in empty flat spacetime, independently of the value of the Lagrangian parameter $\vk$, as long as it real and non-zero, 
and hence without any fine tuning of $\vk$ or any other parameters of the Lagrangian. 

The lowest energy vacuum solution of zero `electric' field strength $F^{0123}\!=\!0$ is automatically consistent with 
a flat spacetime solution of Einstein's eqs.~and consistent also with flat spacetime being the ground state of gravity 
in the complete absence of sources or boundaries. Thus the recasting of the cosmological term in terms of a $4$-form field 
strength as opposed to treating $\La$ as a fixed parameter of the classical theory not only admits a flat space solution of Einstein's 
eqs.~for any value of the still arbitrary parameter $\vk$, without any fine tuning, but it also requires the otherwise 
free integration constant $\tF = 0$ to vanish in empty flat space. This removes one oft-stated obstacle to the solution 
of the `cosmological constant' problem \cite{WeinbergRMP:1989}, and provides a definite stable ground state of energy for the
gravitational vacuum in flat space, against which any excitation can be measured. 

As a classical field strength, the condition (\ref{Einflat}) of $F^{0123} = 0$ in flat empty space is not sensitive to or 
renormalized by extreme UV or Planck-scale physics, any more than a measurable classical background electric field 
is in electromagnetism. The macroscopic boundary condition (\ref{Einflat}) remains valid in the low energy EFT of gravity, 
when all quantum matter effects are included. On the other hand the coupling parameter $\vk$ becomes a scale dependent 
running coupling analogous to $e^2_{\rm eff} (k^2)$ in $D=2,4$ dimensions when quantum effects are included, 
{\it c.f.} \cite{Seiler:2002} and Sec.~\ref{Sec:Tors}.

\section{The Chern-Simons $3$-Form and Massless Fermion Coupling to Torsion}
\label{Sec:Tors}

The general differential geometry of four dimensional spacetime provides a natural candidate for the $4$-form field strength of Sec.~\ref{Sec:F},
namely the $4$-form of the Euler class in $D=4$,
\vspace{-4mm}
\be 
\mathsf{F}\equiv \e_{abcd}\, \mathsf{R}^{ab}\wedge  \mathsf{R}^{cd} = \sdfrac{1}{4} \e_{abcd}\, R^{ab}_{\ \ \,\a\b}\, R^{cd}_{\ \ \,\g\l}
\, dx^\a \wedge dx^\b \wedge dx^\g \wedge dx^\l
\label{Fdef}
\vspace{-4mm}
\ee
which is defined in terms of the curvature $2$-form 
\vspace{-4mm}
\be
\mathsf{R}^{ab} = d\w^{ab} +\w^{ac} \wedge \w^{\ \,b}_c \equiv \sdfrac{1}{2} R^{ab}\!_{\,\m\n}\, dx^\m\wedge dx^\n
\label{Car2}
\vspace{-5mm}
\ee
in which 
\vspace*{-3mm}
\be
\w^{ab} = -\w^{ba} = \w^{ab}\!_{\,\m}\, dx^\m
\label{affcon}
\vspace{-2mm}
\ee
is the affine spin connection $1$-form that specifies the law of parallel transport of orthonormal frames 
in tangent space~\cite{EguGilHan:1980}. In (\ref{Fdef})-(\ref{affcon}) the Latin letters $a,b,\dots$ denote 
tangent space indices while the Greek letters $\a,\b,\dots$ denote spacetime manifold coordinate 
indices.

The Euler class $4$-form (\ref{Fdef}) is exact, {\it i.e.}~$\mathsf{F}=\mathsf{dA}$,  where $\mathsf{A}$ 
is the $S\! O(3,1)$ Lorentz frame dependent Chern-Simons $3$-form, defined in terms of $\w^{ab}$ 
by~\cite{PadYal:2011}
\vspace{-3mm}
\be
\mathsf{A}= \e_{abcd}  \left(\w^{ab} \wedge d \w^{cd} + \sdfrac{2}{3}\, \w^{ab} \wedge \w^{ce}\wedge \w^{fd}\, \h_{ef}\right)\,.
\label{eA}
\vspace{-3mm}
\ee
Correspondingly the Hodge $\ast$ dual of $\mathsf{F}$ is the scalar
\vspace{-3mm}
\be
\ast \mathsf{F} = \ast \mathsf{dA}= -\sdfrac{1}{\,3! } \,\na\!_\m \left( \ve^{\a\b\g\m}\mathsf{A}_{\a\b\g}\right) 
\stackrel{\rm Rie}{=} -E 
\label{EdA}
\vspace{-3mm}
\ee
which is a total divergence, {\it cf.} Appendix \ref{App:CurvStatSph}. It follows  by Stokes' theorem that 
the $4$-volume integral $\int \mathsf{F}  = -\int \!d^4\!x\sqrt{-g}\,  E$ is a pure $3$-boundary term, and 
therefore invariant under all local variations of the bulk $4$-geometry that maintain the $3$-boundary fixed.
Hence it is natural that $\mathsf{F}$ should play a role in the low energy EFT of gravity at macroscopic 
distance scales, independently of short distance physics.

The last equality in (\ref{EdA}) expressing $\ast F$ in terms of the curvature invariant $E$ in (\ref{ECdef}) applies 
in Riemannian geometry in which the affine connection (\ref{affcon}) is the Levi-Civita connection.
In a coordinate basis that connection becomes the familiar Christoffel symbol, expressible entirely in terms of the metric 
and its derivatives, which is symmetric on its last two indices $\G^\m_{\ \a \b} = \G^\m_{\ \b\a}$.

On the other hand the Chern-Simons $3$-form (\ref{eA}) is defined {\it a priori} in terms of the affine spin connection
$\w^{ab}$ {\it independently} of the metric, and is not in general the Levi-Civita/Christoffel connection, in particular
if the geometry admits non-vanishing torsion. The torsion $2$-form~\cite{Cartan:1922,Hehl:1976}
\vspace{-3mm}
\be
\mathsf{T}^a \equiv d\mathsf{e}^a + \w^a\!_{\,b} \wedge \mathsf{e}^b 
=  \sdfrac{1}{2}\, T^a\!_{\,bc}\,\mathsf{e}^b\wedge \mathsf{e}^c
= \sdfrac{1}{2}\, T^a\!_{\,\m\n}\, dx^\m\wedge dx^\n\
\label{Car1}
\vspace{-3mm}
\ee
is defined in terms of $1$-form vielbein or tetrad frame field $\mathsf{e}^a$, also called a soldering form. If $\mathsf{T^a} \neq 0$, 
as it is in a general Einstein-Cartan geometry, the affine connection $\w^{ab}$ contains an anti-symmetric $\G^\m_{\ \a \b} - \G^\m_{\ \b\a} \neq 0$ 
torsional part, and the Chern-Simons $3$-form $\mathsf{A} = \mathsf{A_R}+ \mathsf{A_T}$ in (\ref{eA}) also contains a torsion
dependent contribution $ \mathsf{A_T}$ over and above its Riemannian part $\mathsf{A_R}$. 
In that case of non-vanishing torsion (\ref{EdA}) is modified to
\vspace{-4mm}
\be
\ast \mathsf{F}= \ast d \mathsf{A} = \ast d \big(\mathsf{A_R} + \mathsf{A_T} \big)
= -E + \ast d\mathsf{A_T}
\label{Ftors}
\vspace{-4mm}
\ee
where $E= -(\ast d \mathsf{A_R})$ depends on the metric and Christoffel connection in the absence of torsion
and is given by (\ref{ECdef}), while the additional torsion dependent term in (\ref{Ftors}) is independent of the metric. 

This distinction between $-E$ and the full $\ast\mathsf F$ is significant for the effective action of the conformal anomaly, since
fermions $\Psi$ minimally couple to geometry through the covariant derivative~\cite{Hehl:1976,BirDav,Shap:2002}
\vspace{-5mm}
\be
\na\!_\m\, \Psi=\left(\pa_\m  +\sdfrac{1}{8} \,\w_{ab\,\m}\,  \big[\g^a, \g^b\big] \right)\Psi
\label{Dircov}
\vspace{-2mm}
\ee
intrinsically dependent upon the spin connection $\w^{ab}$, rather than the Levi-Civita/Christoffel connection or the 
metric. In the general setting of Einstein-Cartan geometry allowing for non-zero torsion both
 $\w^{ab}$ and torsion (\ref{Car1}) are defined independently of the metric. Hence the torsion dependent part
$\mathsf{A_T}$ of $\mathsf{A}$, or the full $\mathsf{A}$ itself, generically defined by (\ref{eA}) in terms 
of the spin connection, should be treated as an dynamical variable of low energy gravity that is
{\it a priori} independent of the metric.

If the total $\mathsf{A}$ of the Chern-Simons $3$-form of (\ref{eA}) is identified with the $3$-form potential $A$ of
Sec.~\ref{Sec:F}, {\it i.e.} $\mathsf{A} = A$ and $\mathsf{F} = F$, the massless fermion contribution to the Euler 
term in $\cA$ of (\ref{tranom}) is replaced by $-\tF\, \vf$ according to (\ref{Ftors}), and becomes
\vspace{-5mm}
\be
\begin{aligned}
\big\lag \hat T^\m_{\ \,\m} \big\rag_F &= (a-a_F) \, E - a_F \tF - \frac{2a}{3} \sq R  +  b\, C^2 + \sumi \ \b_i\, \cL_i\\
&\to a' \Big(E - \tfrac{2}{3} \sq R\Big) - a_F \tF +  b\, C^2 + \sumi \ \b_i\, \cL_i\\
&=  - a_F \tF + \frac{\cA}{\!\!\!\sqrt{-g}}\bigg\vert_{a \to a'} \equiv  - a_F \tF +  \frac{\cA'}{\!\!\!\sqrt{-g}}
\end{aligned}
\label{tranomF}
\vspace{-2mm}
\ee
where in the second line, an additional $\sq R$ term from the trivial (non-anomalous) co-cycle has been added, and $\cA'$
is the Riemannian conformal anomaly density (\ref{tranom}) with the massless fermion contribution to the Euler term removed by
the replacement $a \to a' = a- a_F$, {\it i.e.} $\cA \to \cA' = \cA\,\vert_{a\to a'}$.

By repeating the same steps that lead to the local form of the anomaly effective action (\ref{Sanom}) but for (\ref{tranomF})
instead of (\ref{tranom}), one obtains the anomaly effective action
\vspace{-3mm}
\be
\begin{aligned}
&-\frac{a'}{2\ }\!\int\!\!d^4\!x\! \sqrt{-g}\,\bigg\{\big(\sq \vf\big)^2  
- 2\,\Big(R^{\m\n}\! - \!\tfrac{1}{3} Rg^{\m\n}\Big) \,(\pa_\m \vf)\,(\pa_\n\vf)\bigg\}  
+ \frac{1}{2} \!\int\!\!d^4\!x\cA' \,\vf  - \frac{a_F}{2} \int\! d^4\!x \sqrt{-g}\ \tF \,  \vf\\
&\hspace{2cm} = S_{\!\!\cA'} [g;\vf]  + S\!_{\rm int}[A;\vf]
\end{aligned}
\label{SanomF}
\vspace{-3mm}
\ee
where $a' = a- a_F$ is the bosonic contribution to the $a$ anomaly coefficient in (\ref{tranom})-(\ref{bbprime}), and
\vspace{-3mm}
\be
\hspace{-3mm}S\!_{\rm int}[A;\vf] =\! -\frac{a_F}{2}\! \int\! d^4\!x \sqrt{-g}\ \tF \,  \vf 
= -\frac{a_F }{2} \frac{1}{\,3! } \!\int   d^4\!x \sqrt {-g}\,  \ve^{\a\b\g \m}\, A_{\a\b\g}\, \pa_\m \vf
=  \frac{1}{\,3! } \!\int \!  d^4\!x \sqrt {-g}\,   J^{\a\b\g} A_{\a\b\g}
\label{Sint}
\vspace{-2mm}
\ee
after making use of the fact that $\sqrt{-g}\,\tF$ is a total derivative, integrating by parts, and discarding the surface term.
The $3$-form current in (\ref{Sint}) is
\vspace{-4mm}
\be
J^{\a\b\g} =  -\frac{a_F}{\! 2} \, \ve^{\a\b\g \m}\, \pa_\m \vf\,,\quad\qquad 
a_F = -\frac{\hbar}{(4 \pi)^2\!}\, \frac{11}{360}\,  N_F 
\label{Jphi}
\vspace{-4mm}
\ee
for $N_F$ massless Dirac fermions.

Thus the Chern-Simons $3$-form potential $\mathsf{A}$ (\ref{eA}) with field strength $\mathsf{F}$ (\ref{Fdef}),
which appears in the conformal anomaly for massless fermions, provides a $\int\! J\cdot A$ interaction term 
in the effective action analogous to that of ordinary electromagnetism. Despite first appearances $S\!_{\rm int}$ 
in (\ref{Sint}) is in fact independent of the metric, since $\sqrt{-g}\, \ve^{\a\b\g\m} =- [\a\b\g\m]$, defined by 
(\ref{epsdens})-(\ref{epsup}) is a pure number $(0, \pm 1)$, and both $\pa_\m \vf$ and $A_{\a\b\g}$ with 
all lower indices are independent of $g_{\m\n}$ with $A = \mathsf{A}$ defined by (\ref{eA})
in terms of the general spin connection $\w^{ab}$ independently of the metric.

Taking account of the action (\ref{Maxw}), and with the identification $\mathsf{F} \!=\! F \!=\! d A$, the 
total variation with respect to the gauge potential $A$ independently of the metric results in the `Maxwell' eq.
\vspace{-3mm}
\be
\na_{\!\l} F^{\a\b\g\l} = \vk^4  J^{\a\b\g} = -\frac{\vk^4a_F}{\! 2}\, \ve^{\a\b\g\l}\,\pa_\l\vf \qquad {\rm or} \qquad 
\pa_\m \tF =  \frac{\vk^4a_F}{\! 2}\, \pa_\m\vf 
\label{Maxeq}
\vspace{-3mm}
\ee
for its dual. As in ordinary electromagnetism, the conservation of the current (\ref{Jphi}) follows from the identity 
\vspace{-3mm}
\be
d \ast F = \vk^4 \ast J \quad \Longrightarrow \quad d^2\! \ast F = d \ast J  = 0
\label{consJ}
\vspace{-2mm}
\ee
expressed in the language of differential forms, here equivalent to $\pa_{[\m}\pa_{\n]}\,\vf =0$.
Current conservation follows from the local gauge invariance enjoyed by (\ref{Maxw}) and (\ref{Sint}) under the $U(1)$ gauge 
transformation $A\! \to \!A + d \Y$, where the phase angle $\Y \!=\! \e_{abcd}\, \th^{ab} d\w^{cd}$ is a 
$2$-form abelian projection of the $S\!O(3,1)$ group of Lorentz transformations with 
the $6$ parameters $\th^{ab} \!=\! - \th^{ba}$~\cite{EMEFT:2022}. 

The $3$-form current (\ref{Jphi}) is a conserved current source for (\ref{Maxeq}), which replaces the previous (\ref{FnoJ}), 
when massless fermions are present to contribute to the anomaly.  Since the effective vacuum energy $\La_{\rm eff}$ is given 
by (\ref{Lameff}) in terms of $\tF$, the immediate consequence of (\ref{Maxeq}) is that $\La_{\rm eff}$ will change if and when 
$\vf$ does, whenever $a_F \neq 0$ and massless fermions coupling to torsion are present.

\section{Light Fermions and Topological Susceptibility in the Near-Horizon Region}
\label{Sec:Suscep}

The relevance of the considerations of the previous two Secs.~\ref{Sec:F}-\ref{Sec:Tors} to vacuum energy at the BH horizon 
is that due to the extreme blueshifting (\ref{Tol}) in the near horizon region, the masses of the lightest fermions can be neglected 
and they can be treated as massless when $\hbar\w_{\rm loc}(r) > m_F c^2$, or when
\vspace{-3mm}
\be
f^{-\frac{1}{2}}= \left(1 - \sdfrac{\rM}{r}\right)^{-\frac{1}{2}} \gtrsim 
\displaystyle \frac{8 \p G M m_F}{\hbar c} =7.52\times 10^9 \ \left(\frac{\! M}{M_{\odot}}\right) 
\left(\frac{m_F c^2}{\rm 0.04\, eV}\right)
\label{neut}
\ee
where the normalization is to the mass scale $m_F$ of the lightest fermion, {\it viz.}~the lightest neutrino~\cite{NeutrinoMass:2021}.

The condition (\ref{neut}) is satisfied at the radial coordinate distance from the horizon
\vspace{-2mm}
\be
|r-\rM| \lesssim  \D r_F  =  \left(\frac{\hbar c}{8 \p G M m_F}\right)^2 \rM = \frac{L_F^2}{4 \rM}
\label{Drtors}
\vspace{-3mm}
\ee
corresponding to the physical distance
\vspace{-3mm}
\be
\hspace{-2mm}L_F\!=\! \int_{\rM}^{\rM +\D r_F} \!\!\!\!dr\, \left(1 - \sdfrac{\rM}{r}\right)^{-\frac{1}{2}}
\!\simeq 2\!\sqrt{\rM\D r_F} = \frac{\hbar}{2 \p m_F c} 
\simeq 7.85 \times 10^{-5}\,\left(\frac{0.04\, {\rm eV}}{m_F c^2}\right) \ {\rm cm}
\label{elltors}
\vspace{-3mm}
\ee
of order of the Compton wavelength of the lightest fermion, independently of $G$ and $M$. 

At distances closer to the horizon than (\ref{elltors}) the $a_F$ anomaly coefficient and the current 
(\ref{Jphi}) become non-zero. By the `Maxwell' eq.~(\ref{Maxeq}) a non-vanishing $3$-current $J^{t\th\f}$ source 
allows--and requires--the previously rigidly constrained $4$-form field, and hence the effective vacuum energy $\La_{\rm eff}$ 
of (\ref{Lameff}) to change where $\vf(r)$ does. The vacuum energy $\La_{\rm eff}$ changes within this boundary layer 
separating the interior de Sitter region with $\r_{_V}\! >\!0$ from the Schwarzschild exterior with $\r_{_V}\!=\!0$,
resulting in the $\mathbb{R}\otimes \mathbb{S}^2$ worldtube topology illustrated in Fig.~\ref{Fig:Tube}, in which 
the edge of the cylindrical region has a physical width of order (\ref{elltors}).

A second consequence of the activation of torsion near the horizon is that the constant $\vk$ becomes a
scale dependent running coupling which characterizes the topological or torsional susceptibility of the gravitational
vacuum. This is analgous to the behavior of the electric coupling $e^2$ in QED$_2$ 
and the topological susceptibility in QCD~\cite{VendiVecc:1980,Seiler:2002}.

In the present case the topological susceptibility and running coupling $\ka_{\rm eff}^4$ follows from the inverse propagator
$\cD^{-1\, \a\b\g}\!_{\m\n\r} $ of the  $3$-form gauge potential $A_{\a\b\g}$. To lowest order this is simply the local contact term 
obtained from the second variation of the classical action $S\!_F$ of (\ref{Maxw}), {\it i.e.}
\vspace{-3mm}
\be
\cD_0^{-1\, \a\b\g}\!_{\m\n\r}(x,x') \equiv - \frac{\d^2 S\!_F} {\d A_{\a\b\g}(x)\, \d A^{\m\n\r}(x')} =
-\frac{\!1}{(3!)^2}\frac{1}{\vk^4}\, \ve^{\a\b\g\l}\,\ve_{\m\n\r\s}\,\na_\l \na^{\prime\,\s}\, \d_g^4(x,x') 
\label{4Dprop0}
\vspace{-3mm}
\ee
where $\vk^4$ is a fixed constant.

When $a_F \neq 0$, the $J\cdot A$ interaction (\ref{Sint}) results in the additional polarization tensor
\vspace{-3mm}
\be
\P^{\a\b\g}\!_{\m\n\r}(x,x') =\frac{\!\!\!-\,i}{(3!)^2} \,\big\lag T J^{\a\b\g}(x) J_{\m\n\r}(x')\big\rag
\label{4Dpol}
\vspace{-3mm}
\ee
where the $3$-form current $J^{\a\b\g}$ is given in terms of $\vf$ by (\ref{Jphi}), and $T$ denotes the time-ordered product.
Making use of (\ref{Jphi}), (\ref{4Dpol}) can be expressed in terms of the tree-level scalar conformalon propagator
\vspace{-4mm}
\be
 i\, \big\lag T\vf(x) \vf(x') \big\rag = -\frac{1}{a'}\, G_4(x,x')
\vspace{-3mm}
\ee
following from $S_{\!\!\cA'}$, where $G_4(x,x')$ is the propagator function for the differential operator $\D_4$, satisfying (\ref{D4del}).
Hence (\ref{4Dpol}) becomes
\vspace{-4mm}
\be
\P^{\a\b\g}\!_{\m\n\r}(x,x') = \frac{\!1}{(3!)^2}\,\frac{a_F^2}{4a'}\, \ve^{\a\b\g\l}\,\ve_{\m\n\r\s}\,\na_\l \na^{\prime\,\s} G_4(x,x') \,,
\vspace{-3mm}
\ee
and the total inverse $3$-form propagator including this vacuum polarization is\pagebreak
\vspace{-4mm}
\be
\begin{aligned}
\cD^{-1\, \a\b\g}\!_{\m\n\r}(x,x')&= \cD_0^{-1\, \a\b\g}\!_{\m\n\r}(x,x') + \P^{\a\b\g}\!_{\m\n\r}(x,x')\\
&\hspace{-1.5cm}=\displaystyle -\frac{\!1}{(3!)^2} \ve^{\a\b\g\l}\,\ve_{\m\n\r\s}\,\na_\l \na^{\prime\,\s}\left(\frac{\d_g^4(x,x')}{\vk^4} 
 - \frac{a_F^2}{4a'}\, G_4(x,x')\right)
\end{aligned}
\vspace{-3mm}
\ee
so that the effective non-local or scale dependent $\vk^{-4}$ is 
\vspace{-3mm}
\be
\vk^{-4}_{\rm eff}(x,x') \equiv \frac{1}{\vk^4}\, \d_g^4(x,x')
 + \frac{a_F^2}{4|a'|}\,G_4(x,x')
\label{kapeffinv}
\vspace{-3mm}
\ee
where the fact that $a' = -|a'| < 0$ has been used. 

Since the $G_4(x,x')$ propagator grows logarithmically large $\!\sim\! \ln (x-x')^2$ at large distances~\cite{AntMazEM:2007}, 
(\ref{kapeffinv}) shows that the topological susceptibity $\big\lag \tF(x)\, \tF(x') \big\rag \propto \vk^4_{\rm eff}(x,x') \to 0$
in this limit, {\it i.e.} the bare or UV susceptibility $\vk^4$ is screened at distance scales much larger than $1/\vk$ 
if $a_F \neq 0$. This is analogous to the complete screening of the electric coupling $e^2$ in QED$_2$ or the topological 
susceptibity in QCD at large distances in the presence of light dynamical fermions~\cite{Seiler:2002}.

The effect of the running coupling $\vk^{-4}_{\rm eff}$ is to replace the classical local action $S\!_F$ (\ref{Maxw}) 
for the $4$-form field $F$ by the non-local effective action
\vspace{-3mm}
\bea
S\!_F \to &&\  \frac{1}{2} \int d^4\!x\! \sqrt{-g}\! \int \!d^4 x'\!  \sqrt{-g'}\,  \tF(x) \,\vk^{-4}_{\rm eff}(x,x')\, \tF(x') \nn
&&\hspace{3mm} = S\!_F + \frac{a_F^2}{8|a'|} \int d^4\!x \!\sqrt{-g} \!\int d^4 x'\!  \sqrt{-g'}\,  \tF(x) \,G_4(x,x')\, \tF(x')
\label{SFfermion}
\vspace{-4mm}
\eea
when $a_F\!\neq\! 0$. Since $\sqrt{-g}\,\tF$ is independent of the metric, including conformal
rescalings, the addition of the conformally invariant non-local term in (\ref{SFfermion}) does not contribute to
the anomaly (\ref{tranomF}).

As in the case of the non-local anomaly action (\ref{Snonl}), the non-locality of (\ref{SFfermion}) can be removed by the introduction 
of an additional scalar field $\j$, similar to $\vf$. With the addition of $S\!_{\rm int}$ in (\ref{SanomF}),
a non-local term of the same form as that in (\ref{SFfermion}) is already generated, but with the opposite sign.  Thus if
\vspace{-3mm}
\be
\begin{aligned}
a' \!\int\!\!d^4\!x\! \sqrt{-g}\,\bigg\{\big(\sq \j\big)^2  &- 2\,\Big(R^{\m\n}\! - \!\tfrac{1}{3} Rg^{\m\n}\Big) \,(\pa_\m \j)\,(\pa_\n\j)\bigg\}  
- a_F\!\int\!\!d^4\!x\! \sqrt{-g}\,\tF\,\j \\
&= -2\,  S\!_{\cA'}^{(2)}[g;\j] + 2\, S\!_{\rm int} [A; \j] 
\end{aligned}
\label{Spsiloc}
\vspace{-2mm}
\ee
is added to the effective action of (\ref{SanomF}), the total non-local $\int_x\int_{x'}\tF(x) \,G_4(x,x')\, \tF(x')$ term of (\ref{SFfermion}) is reproduced 
with the correct sign and coefficient. The net result is that when there are massless fermions present, the quantum anomaly and scale 
dependent topological susceptibility of the gravitational vacuum (\ref{kapeffinv}) are taken into account by the effective action
\vspace{-4mm}
\be
S\!_F[g;A] + S_{\!\!\cA'} [g;\vf]  -2\,  S\!_{\cA'}^{(2)}[g;\j]  + S\!_{\rm int}[A;\vf + 2\j]
\label{SFloc}
\vspace{-5mm}
\ee
where $S\!_{\cA'}^{(2)}[g;\j]$ is the part of $S\!_{\cA'}[g;\j]$ that is quadratic in $\j$, as in (\ref{Spsiloc}). 

\section{Effective Action and Variational Equations in the Near-Horizon Region}
\label{Sec:EffAct}

Collecting the results of the previous sections, the effective action for gravity requires distinguishing two different cases
depending upon whether or not there are fermions that are sufficiently light to be treated as massless and hence contributing
to the conformal anomaly. Far outside the near horizon region, assuming no strictly massless fermions in the SM, $a_F = 0$ 
and the effective action is
\vspace{-3mm}
\be
S\!_{\rm eff }^{\,\rm (I)}[g,A;\vf] = S\!_{EH}[g] + S\!_F[g;A] + S\!_\cA [g;\vf]\,,\qquad\qquad {\rm Case\, (I)\!:}\ \  a_F = 0
\label{Seff1}
\vspace{-3mm}
\ee
at least at energy scales much less than the mass of the lightest fermions of the SM. In (\ref{Seff1})
$S\!_{EH}[g] = (1/16 \p G)\! \int\! d^4 x \sqrt{-g}\, R $ is the Einstein-Hilbert action of classical GR, $S\!_\cA [g,\vf]$ 
is the anomaly effective action (\ref{Sanom}) and $S\!_F[A]$ is the free `Maxwell' action (\ref{Maxw}) of the $4$-form 
gauge field $F=dA$.

In this Case (I) with $a_F\!=\!0$ and  $J^{\a\b\g}\! =\! 0$, which applies in both the de Sitter interior and Schwarzschild exterior 
regions, (\ref{FnoJ}) implies that $\tF$ is constant, equivalent to a cosmological constant term.
However, since these interior and exterior regions are separated by a thin boundary layer where $J^{\a\b\g}\neq 0$ 
and the value of $\tF$ and $\La_{\rm eff}$ changes according to (\ref{Maxeq}) and (\ref{Lameff}), the effective 
cosmological `constant'-- more properly the local vacuum energy density -- is different in each region. 
In the de Sitter interior region $r < \rM$ and $\rM -r > \D r_F$
\vspace{-5mm}
\be
\La_{\rm eff,\, dS} = 3H^2 = \frac{3}{\rM^2}\,, \qquad \tF\!_{dS} =
\left(\frac{3}{4 \p}\right)^{\!\frac{1}{2}} \frac{\vk^2 M_{\rm Pl}}{\rM}
\label{LamdS}
\vspace{-3mm}
\ee
according to (\ref{Lameff}). On the other hand
\vspace{-5mm}
\be
\La_{\rm eff,\, S} = 0\,,\qquad \tF\!_S = 0
\label{LamS}
\vspace{-4mm}
\ee
in the Schwarzschild exterior region $r > \rM$ and $r-\rM > \D r_F$. In each of these regions well removed from the horizon at 
$r=\rM =H^{-1}$, $\tF$ is constant and the stress tensor $T\!_\cA^{\m\n}$ following from the anomaly action $S\!_\cA [g;\vf]$ 
is of order $1/(\rM^4 f^2)$ with $f\! \sim \!1$, parametrically smaller by $(L_{\rm Pl}/\rM)^2 \!=\! \e^2 \! \sim \!10^{-77}$ 
than the classical terms, and hence quite negligible. Thus the geometry well outside the horizon region is very well 
approximated by the classical interior de Sitter and exterior Schwarzschild solutions of (\ref{fhgrav}).

On the other hand in the near-horizon region defined by (\ref{Drtors})-(\ref{elltors}), $a_F \neq 0$, so that taking account 
of the replacements (\ref{SanomF}) and (\ref{SFloc}) which apply in this region, the gravitational effective action is
\vspace{-3mm}
\be
\begin{aligned}
S\!_{\rm eff}^{\,\rm (II)}[g,A;\vf,\j] =\  & S\!_{EH}[g]  + S\!_F [g;A] + S\!_{\cA'}[g; \vf] - 2\,S\!_{\cA'}^{(2)} [g;\j]  + S\!_{\rm int}[A;\vf+2\j] 
\\[-.2ex]  &{\rm Case\, (II)\!:}\ \  a_F \neq 0
\label{Seff2}
\end{aligned}
\vspace{-4mm}
\ee
instead of (\ref{Seff1}).

The switching from (\ref{Seff1}) to (\ref{Seff2}) when $\hbar\w_{\rm loc}(r) = m_F c^2$ is certainly a overly simplified approximation 
to the actual behavior, as this is a threshold effect where local energies gradually approach and then exceed the fixed energy 
scale $m_F c^2$, so that $a_F$ grows from zero to the finite value (\ref{Jphi}) continuously as $|r-\rM| \to 0$. As in other EFT's,
it is neverheless a reasonable first approximation to treat the threshold effects of one of more mass scales satisfying 
$\hbar\w_{\rm loc}(r) = m_F c^2$ as discontinuous steps with appropriate matching conditions, to guarantee a continuous 
matching of the EFT's below and above the new mass scale. This simplified approach will be adopted here, with
the matching conditions required imposed by boundary conditions on the $\vf$ and $\j$ fields so that the geometries
across the interfaces between (\ref{Seff1}) and (\ref{Seff2}) at $r= \rM \pm \D r_F$ are matched as smoothly as possible,
with no discontinuity in the surface gravities (\ref{surfgrav}) there. This procedure would have to be repeated at each
energy scale at which additional light fermion species can be regarded as massless and contributing to the anomaly, changing $a_F$.
For simplicity only one such fermion scale will be treated here explicitly, with the generalization to including additional
fermions straightforward.

The Euler-Lagrange variational eqs.~that follow from the effective action (\ref{Seff2}) in Case (II) are:
\begin{enumerate}[label=(\arabic*), topsep=0pt,itemsep=-6pt,partopsep=4pt, parsep=4pt]
\item The `Maxwell' eq., $d \ast F= \ast J$ obtained by variation with respect to $A_{\a\b\g}$, the dual of which is
\vspace{-4mm}
\be
\pa_\m \tF = \frac{a_F\, \vk^4}{2}\, \pa_\m\,\big(\vf + 2 \j\big)\,;
\label{Feqloc}
\vspace{-4mm}
\ee
\item The linear eqs.~for $\vf$ and $\j$, 
\vspace{-3mm}
\be
\begin{aligned}
&\D_4 \vf =  \frac{E}{2} -\frac{\sq R}{3} -\frac{a_F}{2a'}\, \tF + \frac{1}{2a'} \left(b\, C^2 +  \sumi \ \b_i\, \cL_i\right)\\[1ex]
&\hspace{1.5cm}\D_4\j =  + \frac{a_F}{2a'}\,  \tF 
\end{aligned}
\label{phipsieqs}
\vspace{-3mm}
\ee
obtained by variation with respect to $\vf$ and $\j$ respectively; 

\item The semi-classical Einstein eqs.,
\vspace{-3mm}
\be
R^{\m\n} - \sdfrac{1}{2} R g^{\m\n} + \La_{\rm eff} g^{\m\n} = 
8 \p G \, \bigg(T\!_{\cA'}^{\,\m\n}[g;\vf] - 2\, T\!_{\cA'}^{(2)\,\m\n}[g;\j]  \bigg)
\label{scEin}
\vspace{-3mm}
\ee
\end{enumerate}
in which $\La_{\rm eff}$ is given by (\ref{Lameff}) and the tensors  $T\!_{\cA'}^{\m\n}, T\!_{\cA'}^{(2)\,\m\n}$ 
obtained by variation of the two $S\!_{\cA'}$ terms in (\ref{Seff2}) which are given in Appendix \ref{App:TAnom}.  
Since $S\!_{\rm int}[A;\vf]$ is independent of the metric by (\ref{epsup}), it makes no contribution to the Einstein eqs.~(\ref{scEin}). 

When $a_F\!=\!0$, eqs.~(\ref{Feqloc})-(\ref{scEin}) reduce to those from variation of the effective action (\ref{Seff1}) 
of Case (I), solved by (\ref{LamdS}) and (\ref{LamS}) together with (\ref{phidS}) and (\ref{phiS}) in the classical 
de Sitter and Schwarzschild regions respectively, and the anomaly stress tensor terms are negligibly small. See however \cite{EMSGW:2017}.
 
For the $S\!_F$ and $S\!_{\rm int}$ terms in the effective action involving $A$, upon noting the gauge freedom 
$A \to A + {\mathsf d}\Y$ where $\Y$ is a co-exact $2$-form, there is in fact only one gauge invariant field variable 
in the $4$ independent components of $A_{\a\b\g}$. This can be chosen to be the component $A_{t\th\f}$ which can be expressed
\vspace{-6mm}
\be
A_{t\th\f} = W(r) \,  \sin\th 
\label{AWdef}
\vspace{-2mm}
\ee
in the coordinates (\ref{sphsta}), so that both $W$ and
\vspace{-4mm}
\be
\tF = \frac{1}{\!\sqrt{-g}}\, \frac{\pa A_{t\th\f} }{\pa r}  = \frac{1}{r^2} \sqrt{\frac{h}{f}}\,  \frac{d W}{dr}
\label{FdW}
\vspace{-3mm}
\ee
are functions only of $r$.  The potential function $W(r)$ is determined in terms of $\tF$ by (\ref{FdW})
up to a gauge dependent constant. In view of (\ref{fhgrav}), (\ref{LamdS}) and (\ref{FdW}) one may take 
\vspace{-2mm}
\be
W_{dS}(r) = \frac{r^3}{\!6}\, \tF\!_{dS} + const.= \frac{r^3}{\!\!\sqrt{3\p}}\,  \frac{\vk^2\, M_{\rm Pl}}{4\, \rM}+ const.
\label{WdS}
\vspace{-2mm}
\ee
in the de Sitter region, where the integration constant is gauge dependent and of no importance.

\section{Boundary Layer Theory and the Rescaled Effective Action in the Layer}
\label{Sec:Rescale}

The essential idea of boundary layer theory~\cite{BendOrzag} is to exploit the existence of the small parameter $\e\! \ll\!1$ multiplying
higher derivative terms in differential eqs., by dividing the domain of the independent variable ($r$ here) into distinct regions. 
There are the two regions outside the boundary layer where the higher order terms in $\e\!\propto\!\!\sqrt{\hbar}$ are negligibly small,
and the classical solution of Sec.~\ref{Sec:gravcl} is known for $r < \rM$ and $r > \rM$. These are separated by a very thin boundary 
layer around $r \simeq \rM$ where the higher derivative quantum terms cannot be neglected, but where both the $r$ coordinate is rescaled 
\vspace{-4mm}
\be
r = \rM\,  ( 1 + \e\,   x)\,, \qquad dr = \e \, \rM
\label{xdef}
\vspace{-4mm}
\ee
together with the metric functions
\vspace{-3mm}
\be
\begin{aligned}
f(r) &= \e\, \mf(x)\\
h(r) &= \e\, \mh(x) 
\end{aligned}
\label{mfmh}
\vspace{-1mm}
\ee
which tend to zero at $r=\rM$ in the classical solution (\ref{fhgrav}). This would produce a singular metric, if not for
the higher derivative quantum terms in the boundary layer around $x=0$. There in the effective action and variational Euler-Lagrange 
eqs.~in $x$ that result from it, only the leading order terms in $\e \ll 1$ are retained, with $\mf$ and $\mh$ treated as order
$\e^0$, which allows for some simplification. When these eqs.~in $x$ are solved, the asymptotic form of the resulting solution for
\vspace{-4mm}
\be
|x| \gg 1 \qquad {\rm but} \qquad \e\, |x| =\Big |\frac{\! r}{\rM}- 1\Big| \ll 1 
\label{matchx}
\vspace{-3mm}
\ee
is matched to that of the classical de Sitter and Schwarzschild solutions as $r \to \rM$ on either side of the near-horizon boundary 
layer. 

This matching of the asymptotic behavior of the boundary layer solution to the classical de Sitter and Schwarzschild solutions 
is possible in the overlap region where each of the inequalities in (\ref{matchx}) can be satisfied, for example when 
$|x| \sim 1/\!\!\sqrt{\e}\ $ for $\e \ll 1$. The result is a global solution to the matching of interior to exterior classical solutions 
through the thin quantum boundary layer that is valid to leading order in $\e$ as $\e \to 0$. This is expected to be an excellent 
approximation for macroscopically large compact objects, given the extreme smallness of $\e= L_{\rm Pl}/\rM$ in (\ref{epsdef}).

For the classical Einstein-Hilbert term, in the metric of (\ref{sphsta}), we have from (\ref{EH}), 
\vspace{-2mm}
\be
S\!_{EH} \to \frac{\left[ 4\p\!\int\! dt\right]}{16 \p G}  \!\int \! dr\!   \sqrt{\sdfrac{f}{h}}\, \left(1-h  - r\frac{dh}{dr}\right) 
\to \frac{ \left[4\p \rM \e\! \int\! dt\right] }{8 \p G}\int \! dx \sqrt{\sdfrac{\mf}{\mh}}\, \left(1 - \frac{d\mh}{dx}\right) 
\label{EHx}
\vspace{-1mm}
\ee
after substituting (\ref{xdef})-(\ref{mfmh}), integrating over angles, and neglecting terms higher order in $\e$.
The overall factor $\left[4\p \rM \e\! \int\! dt\right] /8\p G$ can be extracted from each of the succeeding terms in the
effective action (\ref{Seff2}) in order to compare them to (\ref{EHx}). 

For the $S\!_F$ and $S\!_{\rm int}$ terms in (\ref{Seff2}) involving $A$, $\tF\!_{\rm dS}$ of (\ref{LamdS}) sets 
the scale for $\tF$ and therefore $W$ in the boundary layer, necessitating $W \sim \vk^2\, M_{\rm Pl}\, \rM^2\, \e$.
Defining the precise rescaling of $W$ and $\tF$ by 
\vspace{-3mm}
\be
W(r) = \sqrt{\!\frac{3}{16 \p G}} \ \vk^2 \rM^2\, \e\, w(x)\,,\qquad 
\tF \to \sqrt{\!\frac{3}{16 \p G}} \ \frac{\vk^2}{\rM} \sqrt{\frac{\mh}{\mf}}\ \frac{dw}{dx}
\label{WFwscale}
\vspace{-3mm}
\ee
gives 
\vspace{-3mm}
\be
\begin{aligned}
S\!_F[g;A]  & \to \frac{\textstyle\left[4\p \rM \e\! \int\! dt\right]}{8 \p G}\ \frac{3}{4}
\int \!dx\!  \sqrt{\sdfrac{\mh}{\mf}} \, \left(\frac{d w}{dx}\right)^2 \\[1.5ex]
S\!_{\rm int}[A;\vf] & \to \frac{\left[ 4\p \rM \e \textstyle{ \int} dt\right]}{8\p G}\ 
\g \int\! dx \,\frac{d w}{dx}\ \vf \,.
\end{aligned}
\label{SFsolnx}
\vspace{-1mm}
\ee
where 
\be
\g \equiv -\!\sqrt{3\p G}\, a_F \,\vk^2 \rM = \frac{\!\!\sqrt{3\p}}{4} \, |a_F|\, (\vk \ell)^2 
=  \frac{\!\!\!\sqrt{3\p}}{\, (4\p)^2} \, \frac{11}{1440}\, N_F\,  (\vk \ell)^2
\label{gamdef}
\ee
is a dimensionless parameter depending on $\vk$, and $\ell = 2 \!\sqrt{L_{\rm Pl}\rM}$ is defined in (\ref{elldef}).

Referring to Appendices \ref{App:CurvStatSph} and \ref{App:Rescale}, the term in the anomaly effective 
action (\ref{Seff2}) quadratic in $\vf$ is
\vspace{-3mm}
\begin{align}
&\ S\!_{\cA'}^{(2)}[g;\vf] = \frac{a'}{2\,}\!\int\!\!d^4\!x\sqrt{-g}\,\bigg\{\!-\big(\sq \vf\big)^2  
+ 2\,\Big(R^{\m\n}\! - \!\tfrac{1}{3} Rg^{\m\n}\Big) \,(\pa_\m \vf)\,(\pa_\n\vf)\bigg\}\nn
&  \to\quad \frac{\textstyle\left[4\p \rM \e\! \int\! dt\right]}{8 \p G} \ \left(\frac{- 4 \p Ga'}{\e^2 \rM^2}\right)
\int\! dx  \sqrfr{\mf}{\mh}\left\{ (\D_2 \vf)^2 
-  \frac{\mh \cR}{3} \left(\frac{d\vf}{dx}\right)^2\right\} 
\label{Sanomx}
\end{align}\\[-1.2cm]
to leading order in $\e$, where 
\vspace{-3mm}
\be 
\D_2 \F \equiv \sqrt{\frac{\mh}{\mf}} \frac{d}{dx} \left(\!\sqrt{\mh\mf}\, \frac{d\F}{dx}\right) 
= \mh\, \frac{d^2\F}{dx^2} + \frac{1}{2}\left( \frac{d \mh}{dx} +  \frac{\,\mh}{\mf}\frac{d\mf}{dx} \right)\frac{d \F}{dx}
\label{Delta2}
\vspace{-2mm}
\ee
is the rescaled dimensionless d'Alembert operator of the 2D geometry of (\ref{sphsta}) restricted to the $(t,r)$
hyperplane, operating on scalar functions $\F(r) \to \F(x)$ only, and
\vspace{-3mm}
\be
\cR \equiv \displaystyle -\sqrt{\frac{\mh}{\mf}}\frac{d}{dx} \left(\!\!\sqrt{\frac{\mh}{\mf}} \frac{d\mf}{dx}\right) 
= -\D_2 \ln f
\label{R2}
\vspace{-2mm}
\ee
is the rescaled dimensionless scalar curvature of the 2D geometry of (\ref{sphsta}). The rescaled 
$S\!_{\cA'}^{(2)}[g;\vf, \j]$ is obtained from (\ref{Sanomx}) by substituting $\j$ for $\vf$ and 
multiplying by an overall factor of $-2$.

Making use of (\ref{WeylErescale}), the Weyl squared $b$ term in (\ref{Seff2}) linear in $\vf$ is
\vspace{-3mm}
\be
\frac{b}{2} \int \! d^4 x\sqrt{-g}\, C^2 \,\vf 
\to   \frac{\textstyle\left[4\p \rM \e\! \int\! dt\right]}{8 \p G} \ \left(\frac{4 \p G b}{3\e^2 \rM^2}\right)
\int \! dx   \sqrt{\sdfrac{\mf}{\mh}}\  \cR^2\,\vf
\label{Csqx}
\vspace{-2mm}
\ee
to leading order in $\e$. Similarly utilizing (\ref{sqRrescale}), the $\sq R\, \vf$  term in (\ref{Seff2}) becomes
\vspace{-3mm}
\be
\frac{a'}{\!2} \int \! d^4 x\sqrt{-g}\ \left(- \sdfrac{2}{3}\sq R \,\vf \right)
\to  \frac{\textstyle\left[4\p \rM \e\! \int\! dt\right]}{8 \p G}\ \left(\frac{-8\p Ga'}{3\e^2\rM^2}\right)
\int \! dx   \sqrt{\frac{\mf}{\mh}}\,  \cR\, \D_2 \vf 
\label{RsqPhix}
\vspace{-2mm}
\ee
to leading order in $\e$ in the boundary layer, after integration by parts and discarding of surface terms.

In contrast to the $C^2$ and $\sq R$ integrands which are each of order $1/\e^2$, from (\ref{WeylErescale}) $E$ is of order $1/\e$
and hence the Riemannian Euler term in (\ref{Sefflay}) is suppressed by a factor of $\e$ relative to (\ref{Csqx}) and (\ref{RsqPhix}), 
and can be dropped to leading order in $\e \ll 1$. This insensitivity to $E$ in (\ref{tranomF}) shows that the effective action in the
boundary layer is independent of whether one identifies only the torsional part of the Chern-Simons potential 
$\mathsf{A}_T$ with the $3$-form potential $A$ of Sec.~\ref{Sec:F} or the full $\mathsf{A} = \mathsf{A}_R + \mathsf{A}_T = A$.

The $\sum_i \b_i\cL_i$ terms in the anomaly effective action can also be neglected to leading order in $\e$, 
provided that $\cL_i$ contain only up to second derivatives of the matter fields, as in the SM, and there are no other rescalings 
of the gauge or matter fields to be accounted for in the boundary layer.

Using the results (\ref{EHx}), (\ref{SFsolnx}), (\ref{Sanomx}), and (\ref{Csqx})-(\ref{RsqPhix}), the rescaled effective action 
for the functions $(\mf, \mh, \vf,\j, w)$ to be extremized in the near-horizon region becomes 
\vspace{-3mm}
\begin{align}
&\cS[\mf,\mh;\vf, \j, w] = \int \! dx \sqrfr{\mf}{\mh}\, \left( 1 - \frac{d\mh}{dx} \right)
\ +\frac{3}{4} \!\int \!dx  \sqrt{\sdfrac{\mh}{\mf}} \, \left(\frac{d w}{dx}\right)^2
+ \g\!\int\! dx \,\frac{d w}{dx}\ \big(\vf + 2\,\j\big) \nn
&\hspace{8mm}+ \a \!\int \! dx \sqrfr{\mf}{\mh}\,  
\left\{ \big(\D_2 \vf\big)^2  -  \frac{\mh \cR}{3} \left(\frac{d\vf}{dx}\right)^2  + \frac{2}{3}\, \cR\,  \D_2\vf \right\} 
+ \b \!\int \! dx   \sqrfr{\mf}{\mh}\ \cR^2\, \vf \\
&\hspace{3cm}-2 \a\!\int \! dx \sqrfr{\mf}{\mh}\,  
\left\{\big(\D_2 \j\big)^2  -  \frac{\mh \cR}{3} \left(\frac{d\j}{dx}\right)^2\right\} 
\label{Sefflay}
\end{align}\\[-1cm]
after removing an overall factor of $\left[ 4\p \rM \e \!\int\! dt\right]/8\p G$, and defining the dimensionless coefficients
\vspace{-2mm}
\begin{align} 
\a &\equiv -\frac{4 \p G a'}{\e^2 \rM^2} = \frac{4 \p\, |a'|}{\hbar} \left(\frac{L_{\rm Pl}}{\e \rM}\right)^2 = 
\frac{1}{4 \p}\, \frac{1}{360}\ \Big(N_S + 62 N_V\Big)\\
\b &\equiv \frac{4\p G b}{3\e^2 \rM^2} = \frac{4\p b}{3\hbar} \left(\frac{L_{\rm Pl}}{\e \rM}\right)^2=
\frac{1}{4 \p}\,\frac{1}{360}\  \Big(N_S + 6 N_F + 12 N_V\Big)\,.
\label{ab}
\end{align}
\vspace{-7mm}

\begin{wrapfigure}{r}{.42\textwidth}
\vspace{-5mm}
\hspace{3mm}
\includegraphics[height=9cm,width=8cm, trim=5cm 9.4cm 4.5cm 9cm, clip]{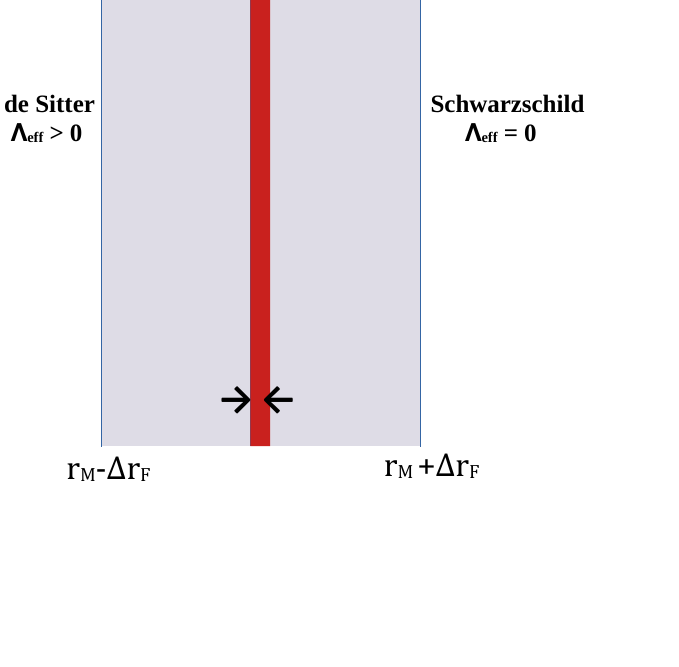}
\vspace{-3.35cm}
\singlespace
\caption{Expanded view of the boundary layer at $r\!\simeq\!\rM$, showing the central region of 
$\D r\! \simeq\! L_{\rm Pl}$ (red, denoted with arrows), enclosed by the wider layer with $\D r_F$ 
of (\ref{Drtors}) (blue violet), where the lightest fermion can be treated as massless and $\La_{\rm eff}$ 
varies.}
\label{Fig:Layer}
\vspace{-1cm}
\end{wrapfigure} 

The $1/\e^2$ scaling of the $\a$ and $\b$ higher derivative terms in (\ref{Sefflay})-(\ref{ab}) allows them
to compensate for their very small coefficient $\hbar G/\rM^2 \!=\!(L_{\rm Pl}/\rM)^2$. Since $\a$ and $\b$ are of order $1$, 
the anomaly stress tensor significantly affects the classical geometry in the boundary layer region where $x$ is of order $1$ 
and $|r-\rM|$ is of order $L_{\rm Pl}$, corresponding to the physical thickness of the layer $\ell = 2 \sqrt{L_{\rm Pl} \rM}$ 
of (\ref{elldef}), and consistent with the behavior of (\ref{ThorizM}) and (\ref{ThorizH}). This central region of 
$\D r \simeq L_{\rm Pl}$ is illustrated as the narrow red colored strip in the expanded view of the boundary layer 
in Fig.~\ref{Fig:Layer}.

On the other hand the parameter $\g$ (\ref{gamdef}) depends upon $\vk$ which introduces a scale independent of 
the Planck scale. Since $\g$ becomes non-zero when the lightest fermion(s) of the SM can be treated as massless,
at $\D r_F = L_F^2 /4\rM$, which is much greater than $L_{\rm Pl}$ for typical BH or gravastar masses,  (\ref{Drtors})
defines a wider boundary layer, illustrated as the light blue violet shaded region in Fig.~\ref{Fig:Layer}. 
In this wider but still small range of physical distances from $\rM$, $\tF$ and $\La_{\rm eff}$ vacuum energy vary
according to (\ref{Feqloc}), or
\vspace{-3mm}
\be
\frac{d}{dx}\left(\!\!\sqrt{\frac{\mh}{\mf}} \frac{dw}{dx}\right)= - \frac{2 \g}{3}\,\frac{d}{dx} \big( \vf + 2 \j\big) 
\label{vfsoln}
\vspace{-3mm}
\ee
in terms of the rescaled variables, while $\vf$ and $\j$ satisfy 
\vspace{-3mm}
\begin{align}
\left( \D_2^2 + \frac{\cR}{3} \D_2 + \frac{\mh}{3} \frac{d \cR}{dx}\frac{d}{dx}\right) \vf
&= - \frac{1}{3}\D_2 \cR - \frac{\b}{2\a}  \cR^2 - \frac{\g}{2 \a} \sqrt{\frac{\mh}{\mf}} \frac{dw}{dx} \label{phieq}\\
\left( \D_2^2 + \frac{\cR}{3}\D_2 + \frac{\mh}{3} \frac{d \cR}{dx}\frac{d}{dx}\right) \j 
&= +  \frac{\g}{2\a} \sqrt{\frac{\mh}{\mf}} \frac{dw}{dx} 
\label{psieq}
\nn[-1.4cm]
\end{align}
which are eqs.~(\ref{phipsieqs}) to leading order in $\e$ in the rescaled variables.

\vspace{-1mm}
The Einstein eqs.~(\ref{scEin}) for the rescaled metric functions $\mf, \mh$, following from (\ref{Sefflay}) are
\vspace{-2mm}
\begin{align}
&1 - \frac{d \mh}{dx} -\frac{3\mh}{4\mf} \left(\frac{dw}{dx}\right)^2 \!= \a\,\bigg(\vl_E[\vf] -2\,\vl_E^{(2)}[\j] \bigg) + \b\, \vl_C [\vf]\\
&\frac{\mh}{\mf} \frac{d \mf}{dx} - 1 +\frac{3\mh}{4\mf} \left(\frac{dw}{dx}\right)^2\!
= \a\, \bigg(\zp_E[\vf] -2\,\zp_E^{(2)}[\j] \bigg)+ \b\, \zp_C[\vf] \nn[-1.4cm]
\label{fheqs}
\end{align}
to leading order $\cO(e^0)$ in $\e$ in the boundary layer, where the rescaled energy density and pressure terms in (\ref{fheqs}) are given 
explicitly by (\ref{rhoA})-(\ref{rhopC}) of Appendix \ref{App:TAnom}. 

The two eqs.~(\ref{fheqs}) are the $\m\!=\!\n\! = \!t$ and $\m\!=\!\n\!=\! r$ components of the rescaled semi-classical Einstein eqs., 
which determine the equal $\m\!=\!\n\!=\! \th$ and $\m\!=\!\n\!=\! \f$ transverse pressure components by the covariant conservation  
eq.~$\na\!_\m T^\m_{\ \, r} = 0$ in the case of spherical symmetry.

\section{Boundary Conditions and a Variational Ansatz for the Boundary Layer}
\label{Sec:BoundCond}

At the inner and outer edges of the boundary layer $r= \rM \mp \D r_F$, the solutions for $\vf, \j, \tF$ and the metric functions $f$ and 
$h$ within the layer are required to match those outside of the layer, where $a_F=0$, and the effective action (\ref{Seff2}) reverts to
(\ref{Seff1}) of Case (I). In the rescaled radial coordinate $x$ defined by (\ref{xdef}) these edges of boundary layer occur at
$\e x_{\pm} = \pm \D r_F/\rM = \pm  (L_F/2\rM)^2$ which is of order $10^{-20}$ for typical values of the parameters. One may
therefore make use of the small $|r -\rM |$ limit of the solutions in the classical de Sitter and Schwarzschild regions, neglecting
higher order terms.

For the rescaled metric function $\mf$, this yields the boundary conditions
\vspace{-4mm}
\be
\begin{aligned}
&\mf (x_-)= \frac{|x_-|}{2} = -\frac{x_-}{\!2}\,,\qquad &\mf (x_+)= x_+\\
&\frac{d\mf}{dx}(x_-) = - \frac{1}{2}\,, &\frac{d\mf}{dx}(x_+) = 1
\end{aligned}
\label{fbcs}
\vspace{-3mm}
\ee
while for $\mh$
\vspace{-1mm}
\be
\mh (x_-)= 2\,|x_-| = -2\,x_- \,,\qquad \mh (x_+)= x_+
\label{hbc}
\vspace{-1mm}
\ee
reflecting the different linear slopes of the functions $f$ and $h$ in their approach $r\to \rM$ illustrated in Fig.~\ref{Fig:SchwStar}.
There is no condition {\it a priori} on $d \mh/dx$ in (\ref{hbc}) since eqs.~(\ref{fheqs}) are no higher than third order in derivatives 
of $\mh$, whereas they are generally fourth order in $\mf$ (if $\b \neq 0$). 

Since $\tF= \tF_{dS}$ takes on the de Sitter value (\ref{LamdS}) at $r= \rM -\D r_F$ but $\tF= 0$ at $r= \rM + \D r_F$, 
in the rescaled variables (\ref{WFwscale}) it satisfies the boundary conditions
\vspace{-3mm}
\be
\sqrfr{\mh}{\mf} \frac{dw}{dx}\bigg\vert_{x_-}= 2 \,,\qquad \sqrfr{\mh}{\mf}\frac{dw}{dx}\bigg\vert_{x_+} = 0
\label{wbc}
\vspace{-2mm}
\ee
at $x\!=\!x_\mp$ respectively. Since $\sqrt{\mf/\mh} =2$ at $x=x_-$ from (\ref{fbcs})-(\ref{hbc}), the first condition of (\ref{wbc}) 
is equivalent to $dw/dx = 1$ at $x=x_-$.

The effective action (\ref{Seff2}) must revert to (\ref{Seff1}) in the classical de Sitter and Schwarzshild regions. In order for
$S\!_{\cA'}^{(2)}[g;\j]$ to vanish there, $\j$ is required to be constant and $\j' =0$ in each classical region. This 
requires the boundary conditions
\vspace{-4mm}
\be
\begin{aligned}
&& &\j (x_-)= \j_0 \,,\qquad & \j(x_+)= \j_{\infty}\\
&& &\frac{d \j}{dx}(x_-) =0\,, &  \frac{d \j}{dx}(x_+) = 0
\end{aligned}
\label{psibcs}
\vspace{-3mm}
\ee
on the $\j$ field where $\j_0$ and $\j_{\infty}$ are constants, in order for the solution within the boundry layer to join smoothly to the classical 
regions on each side. 

The values of $\vf$ and $\vf'$ may take on arbitrary values at $x=x_{\pm}$ in order to match to the solutions 
(\ref{phiS}), (\ref{phidS}), and their small $|r-\rM |$ values (\ref{phiSlim}), (\ref{phidSlim}) at the edges of the boundary layer region. 
This gives the boundary conditions
\vspace{-3mm}
\be
\begin{aligned}
&& &\vf (x_-)= c_0 + c_{_H}\ln \left(\frac{\e|x_-|}{2}\right)\,, \qquad &&\vf (x_+)= c_\infty + c_{_S}\ln \left(\e|x_+|\right)\\
&& &\frac{d \vf}{dx}(x_-) = \frac{c_{_H}}{x_-}\,,&&\frac{d \vf}{dx}(x_+) = \frac{c_{_S}}{x_+}
\end{aligned}
\vspace{-2mm}
\label{phibcs}
\ee
for the conformalon $\vf$ field in terms of the four constants $c_0,c_{_H},c_\infty, c_{_S}$ in the classical regions.

A numerical solution of the boundary layer eqs.~(\ref{vfsoln})-(\ref{fheqs}) satisfying these boundary conditions is necessary
in the case of general and realistic values of $\g$ (\ref{gamdef}), which controls the coupling of the $\vf,\j$ fields to $\La_{\rm eff}$
and hence the metric functions $\mf,\mh$ in the layer. When $\g \to 0$, $\tF$ and $\La_{\rm eff}$ become constant and
the effective action (\ref{Seff2}) in the boundary layer reverts to (\ref{Seff1}) in the classical regions. Thus in the case of small $\g$,
$\tF$ and $\La_{\rm eff}$ vary slowly, and the solution of the EFT eqs.~(\ref{vfsoln})-(\ref{fheqs}) can be approximated 
by the classical solution in most of the boundary region excluding the innermost strip of $\D r \sim L_{\rm Pl}$ 
of Fig.~\ref{Fig:Layer}, where the quantum effects become signficant. This suggests the trial ansatz 
\vspace{-4mm}
\be
\begin{aligned}
\mf (x; \h ,\l) &= \frac{1}{2} \sqrt{x^2 + \h^2}\ \T_-(\l x) +  \sqrt{x^2 + \h^2}\  \T_+(\l x)\\
\mh (x; \h, \l) &=2\sqrt{x^2 + \h^2}\ \T_-(\l x) +  \sqrt{x^2 + \h^2} \ \T_+(\l x)\\
\end{aligned}
\vspace{-1mm}
\label{ansfh}
\ee
for the rescaled metric functions $\mf,\mh$, where
\vspace{-4mm}
\be
\T_\pm(\l x) \equiv \sdfrac{1}{2}\, \Big[ 1 \pm {\rm tanh}(\l x) \Big]
\label{Heavi}
\vspace{-4mm}
\ee
which in the limit $\l \to \infty$ become the Heaviside step functions $\T(x)$ and $1 - \T(x)$ respectively, but otherwise allow
for a smooth interpolation between the $x<0$ and $x>0$ portions of the boundary layer, matching to the de Sitter and 
Schwarzschild metrics. The parameter $\h >0$ regularizes the classical metric cusp singularity of Fig.~\ref{Fig:SchwStar} 
at $x=0, r=\rM$, so that if $\h \to 0$ and $\l \to \infty$, (\ref{ansfh}) coincides with the rescaled values of the classical Schwarzschild 
star metric functions (\ref{fhgrav}), by use of (\ref{xdef})-(\ref{mfmh}), to leading order in $\e \ll 1$. Otherwise $\h$ and $\l$ can 
be treated as parameters to be freely varied, to search for an approximate solution to the boundary layer equations.

For $\h \ll1, \l \gg 1$, (\ref{ansfh}) leads also to $\cR = 0$ for the rescaled curvature (\ref{R2}), so that eqs.~(\ref{phieq})-(\ref{psieq})
become 
\vspace{-6mm}
\be
\begin{aligned}
\big(\D_2\big)^2\vf \simeq 4\, \frac{d}{dx} \left[ x \frac{d^2}{dx^2} \left(x \frac{d \vf}{dx}\right)\right]
&\simeq -\frac{\g}{2 \a} \sqrt{\frac{\mh}{\mf}} \frac{dw}{dx}  \\
\big(\D_2\big)^2\j \simeq 4\, \frac{d}{dx} \left[ x \frac{d^2}{dx^2} \left(x \frac{d \j}{dx}\right)\right]
&\simeq +  \frac{\g}{2\a} \sqrt{\frac{\mh}{\mf}} \frac{dw}{dx} 
\end{aligned}
\label{phipsi}
\vspace{-1mm}
\ee
when the definition of $\D_2$, (\ref{Delta2}) for $\mf \simeq -x/2, \mh \simeq -2x$ in the classical approximation for $x<0$ are used as well.
When $\g$ is sufficiently small, the right sides may also be neglected, and one may assume as a lowest order approximation that $\vf$ and $\j$ 
are solutions to the homogeneous versions of (\ref{phipsi}). 

At the same time (\ref{vfsoln}) can be integrated immediately to give
\vspace{-3mm}
\be
\sqrfr{\mh}{\mf} \frac{dw}{dx} = -\frac{2\g}{3}\, \Big( \vf + 2 \j + c_F \Big)
\label{Fvfjsoln}
\vspace{-3mm}
\ee
where $c_F$ is another constant, so that a particular homogeneous solution of (\ref{phipsi}) for the linear combination $\vf + 2 \j$ and
this rescaled $\tF$ is
\vspace{-6mm}
\be
\sqrfr{\mh}{\mf} \frac{dw}{dx} =\left\{\begin{array}{ll} \displaystyle \frac{2 x}{\,x_-\!}\,,\qquad &x_- \le x \le 0 \\[1ex]
\ 0\,, & 0\le x \le x_+ \end{array} \right.
\label{Fans}
\vspace{-4mm}
\ee
which satisfies the boundary conditions (\ref{wbc}), and is monotonically decreasing for $x<0$, with no logarithms, and
with a small slope for $|x_-| \gg 1$. The general homogeneous solutions of the (\ref{phipsi}) that satisfy the boundary 
conditions (\ref{psibcs})-(\ref{phibcs})  and (\ref{Fans}) in the boundary layer are easily found, and
\footnote{A second linearly independent solution $x \ln x $ of the homogeneous eqs.~(\ref{phipsi}) has been omitted 
from (\ref{phi0})-(\ref{psi0}) since it produces large logarithmic terms in the effective action (\ref{Sefflay}), near the endpoints 
$x=x_{\pm}$ of the boundary layer, violating the assumption that this region should be quasi-classical if $\g$ and the variation 
of $\tF$ is small. In contrast, the $\z$ terms in (\ref{phi0})-(\ref{psi0}) give significant contributions to the action and stress 
tensor only in the very narrow central region of the boundary layer where large quantum effects, as required.}
\vspace{-3mm}
\be
\vf (x) = \left\{\begin{array}{ll}-2 \z\,  \displaystyle{\ln\left(\frac{x\,}{x_-\!}\right) 
+ \left(2 \z -\frac{3}{\g}\right) \left( \frac{x\,}{x_-\!}-1\right) + \vf(x_-)}\,,\qquad  & x<0\\[2ex]
 -2 \z\, \displaystyle{\ln\left(\frac{x\,}{x_+\!}\right) + 2 \z \left( \frac{x\,}{x_+\!}-1\right) + \vf(x_+)}\,, & x>0 \end{array}\right.
\label{phi0}
\vspace{-3mm}
\ee
with
\vspace{-3mm}
\be
\j (x) = \left\{\begin{array}{ll}\, \z\, \displaystyle{\ln\left(\frac{x\,}{x_-\!}\right) - \z \left( \frac{x\,}{x_-\!}-1\right) + \j(x_-)}\,,\qquad  & x<0\\[2ex]
\z\, \displaystyle{\ln\left(\frac{x\,}{x_+\!}\right) - \z\left( \frac{x\,}{x_+\!}-1\right) + \j(x_+)}\,, & x>0 \end{array}\right.
\label{psi0}
\ee
may be taken as a trial ansatz for $\vf$ and $\j$, for any value of the free parameter $\z$,  provided
\vspace{-3mm}
\be
\frac{d \vf}{dx}(x_-) = -\frac{3}{\g x_-}\ \Rightarrow\  c_{_H} = - \frac{3}{\g} \,,\qquad\qquad 
\frac{d \vf}{dx}(x_+) =0 \ \Rightarrow c_{_S} =0 
\label{dphimp}
\vspace{-3mm}
\ee
in order to satisfy (\ref{Fvfjsoln})-(\ref{Fans}), and (\ref{phibcs}), and provided that
\vspace{-3mm}
\be
\vf(x_-) + 2\, \j(x_-) + c_F = 0 = \frac{3}{\g} + \vf(x_+) + 2\, \j(x_+) + c_F 
\label{vfjmp}
\vspace{-3mm}
\ee
are also satisfied.
With (\ref{dphimp}) fixing $c_{_H}$ and $c_{_S}$, the remaining two $(c_0, c_\infty)$ of the four integration constants 
of (\ref{phibcs}) in the classical region remain free, and (\ref{vfjmp}) may be viewed as fixing $\j(x_\pm)$.

Regulating all the logarithmic terms in (\ref{phi0})-(\ref{psi0}) by the replacement $|x| \to \sqrt{x^2 + \h^2}$, 
and providing a smooth transition from the $x<0$ to $x>0$ regions by use of the functions $\T_{\pm}$ of (\ref{Heavi}), we
finally arrive at the trial ansatz
\vspace{-3mm} 
\be
\begin{aligned}
\vf(x; \h, \l) &=\left[- \z\,  \ln\left(\frac{x^2 + \h^2}{x_-^2 +\h^2}\right) 
+ \left(2 \z -\frac{3}{\g}\right) \left( \frac{x\,}{x_-\!}-1\right) + \vf(x_-)\right] \,\T_-(\l x) \\
&\hspace{-4mm}+ \left[- \z\, \ln\left(\frac{x^2 + \h^2}{x_+^2 +\h^2}\right) 
+ 2 \z  \left( \frac{x\,}{x_+\!}-1\right) + \vf(x_+) \right] \T_+(\l x)\\[1ex]
\j(x; \h, \l) &=\left[ \frac{\z}{2}\,  \ln\left(\frac{x^2 + \h^2}{x_+^2 +\h^2}\right) 
-\z \left( \frac{x\,}{x_-\!}-1\right) + \j(x_-)\right] \,\T_-(\l x)\\
&\hspace{-4mm}+ \left[\frac{\z}{2}\,  \ln\left(\frac{x^2 + \h^2}{x_+^2 +\h^2}\right) 
-\z \left( \frac{x\,}{x_+\!}-1\right) + \j(x_+)\right] \,\T_+(\l x)
\end{aligned}
\vspace{-2mm}
\label{ansphipsi}
\ee
for $\vf$ and $\j$ in the boundary layer, together with (\ref{Fans}), also made smooth by
\vspace{-4mm}
\be
\sqrfr{\mh}{\mf} \frac{dw}{dx}(x;\l) = \frac{2 x}{\,x_-\!} \,\T_-(\l x)
\label{Fansreg}
\vspace{-2mm}
\ee
consistent with (\ref{Fvfjsoln}). If (\ref{ansfh}) for the metric functions $\mf,\mh$ together with (\ref{ansphipsi})-(\ref{Fansreg})
is substituted into (\ref{Sefflay}) for the effective action of the boundary layer, $\h,\l$ may be treated as variational parameters.
Their values at the extremum of the (\ref{Sefflay}) then provides a simple two parameter variational approximation to the 
solution of (\ref{vfsoln}-(\ref{fheqs}) satisfying the specified boundary conditions. The free parameter $\z$ may also be treated 
as a variational parameter, but in the following $\z =1$ is fixed.

The result of this extremization for $\a= 1, \b=0, \z=1, \e = 10^{-4}, |x_\pm| = 10^2$, and two values of $\g$ is illustrated in Fig.~\ref{Fig:Minima},
showing that a minimum of the effective action (\ref{Sefflay}) for the variational ansatz (\ref{ansfh}), (\ref{ansphipsi})-(\ref{Fansreg})
exists with finite values of $(\h,\l)$ of order one, for both $\g =0.1$ and $\g =1$. 

\begin{figure}
\vspace{-2mm}
\centering
\begin{subfigure}{.47\textwidth}
\centering
\includegraphics[width=\linewidth, height=6.5cm, trim=0cm 0cm 0cm 0cm, clip]{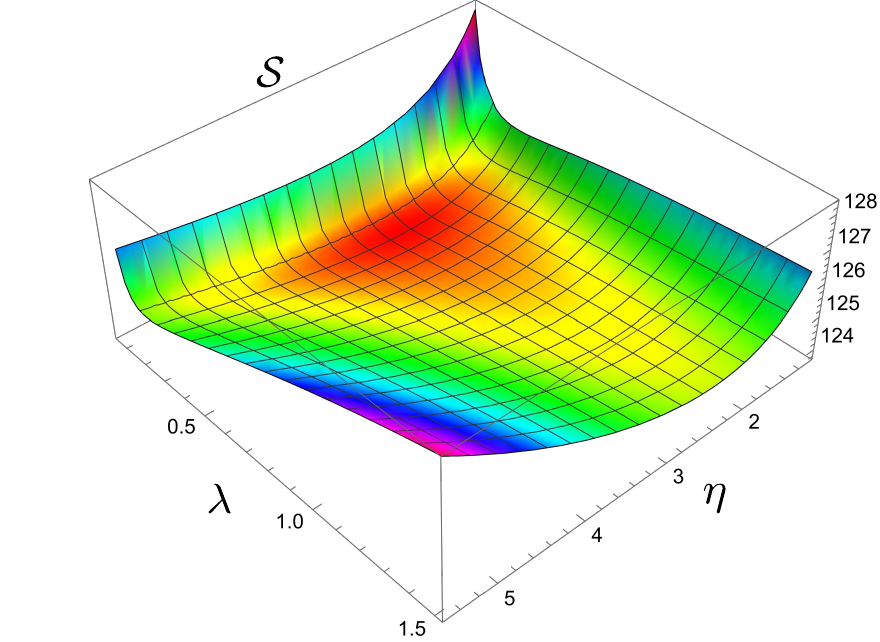}
\vspace{-2mm}
\singlespace
\caption{Minimum of $\cS$ for $\g=0.1$ at $(\h =2.55737,\l = 0.231628)$. \label{Fig:Min0.1}}
\vspace{-2mm}
\end{subfigure}
\hspace{1cm}
\begin{subfigure}{.43\textwidth}
\centering
\includegraphics[width=\linewidth,height=6.5cm, trim=0cm 0cm 0cm 0cm, clip]{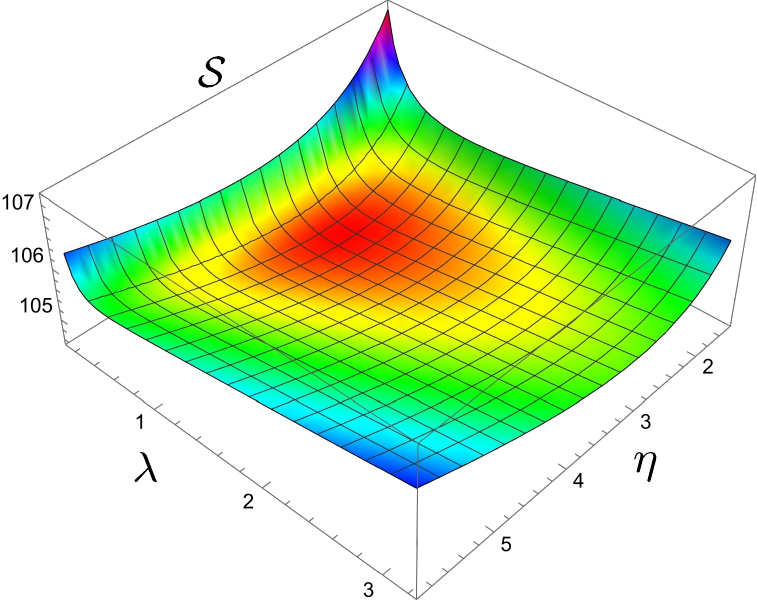}
\vspace{-2mm}
\singlespace
\caption{Minimum of $\cS$ for $\g=1$ at $(\h =2.96653,\l = 0.678154).$\label{Fig:Min1}}
\vspace{-2mm}
\end{subfigure}
\caption{The effective action $\cS$ of (\ref{Sefflay}) for the variational ansatz (\ref{ansfh}), (\ref{ansphipsi})-(\ref{Fansreg}), 
for $\a=1, \b=0$ as a function of $(\h,\l)$ for two values of $\g$, showing a minimum in the deep red region in each case. \label{Fig:Minima}}
\end{figure}

\begin{figure}[h]
\vspace{1mm}
\centering
\begin{subfigure}{.44\textwidth}
\includegraphics[width=\linewidth, height=4.8cm, trim=0cm 0cm 0cm 0cm, clip]{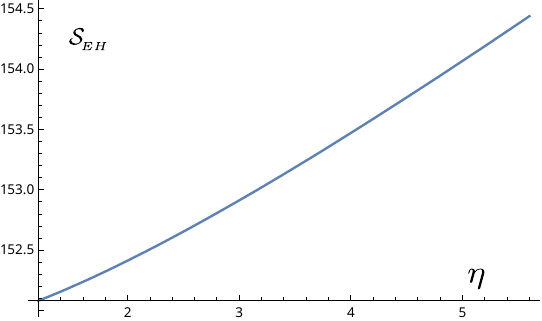}
\singlespace
\caption{Classical Einstein-Hilbert Action $\cS_{_{EH}}$ as function of $\h$ for fixed $\l$. \label{Fig:EHAction(eta)}}
\vspace{-2mm}
\end{subfigure}
\hspace{1cm}
\begin{subfigure}{.44\textwidth}
\centering
\includegraphics[width=\linewidth,height=4.8cm, trim=0cm 1mm 0cm 0cm, clip]{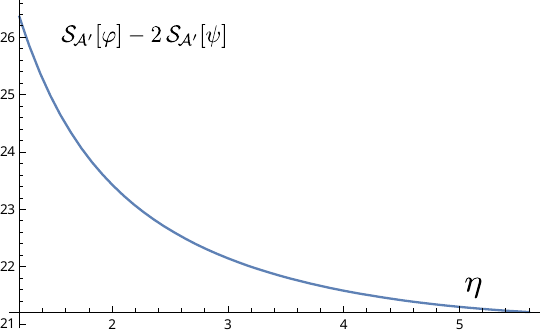}
\singlespace
\caption{Sum of Quantum Anomaly Actions as function of $\h$ for fixed $\l$. \label{Fig:AnomAction(eta)}}
\vspace{-2mm}
\end{subfigure}
\caption{Contrasting behaviors of the classical action and quantum anomaly actions in (\ref{Sefflay}) as functions of $\h$ 
for the variational ansatz  (\ref{ansfh}), (\ref{ansphipsi})-(\ref{Fansreg}), $\g=0.1$ and fixed $\h=2.96653$.  \label{Fig:Seta}}
\vspace{-4mm}
\end{figure}

That a well-defined action minimum exists may be understood by comparing the behavior of the classical Einstein-Hilbert term 
$\cS_{EH}$ and sum of quantum anomaly terms $\cS\!_{\cA'}[\vf] - 2\,\cS\!_{\cA'}[\j]$, as functions of $\h$ and $\l$. Fig.~\ref{Fig:Seta} shows
that the classical EH action increases with increasing $\h$ favoring $\h \to 0$ of the classical solution (\ref{fhgrav}), while 
$\cS\!_{\cA'}[\vf] - 2\,\cS\!_{\cA'}[\j]$ increases in this limit, due to the large higher derivative terms for a narrower central region,
hence favoring larger $\h$ instead. The result of this competition of classical and quantum terms is a balance and minimum of the
total effective action at a finite $\h$.  Likewise Fig.~\ref{Fig:Slam} shows that the classical action favors large $\l \to \infty$, 
corresponding to the sharp Heaviside step function of the the classical solution (\ref{fhgrav}), while this behavior is disfavored by the quantum
terms, again due to the higher derivative terms becoming large as the central region around $x=0, r=\rM$ becomes sharper
and narrower. Again the result is a minimum of the total action at a finite $\l$.  The $\cS_F$ and $\cS_{\rm int}$ terms are independent of $\h$ 
and only weakly dependent upon $\l$, and do not do not greatly influence the balance of classical and quantum terms.

\begin{figure}[h]
\centering
\begin{subfigure}{.44\textwidth}
\includegraphics[width=\linewidth, height=4.8cm, trim=0cm 0cm 0cm 0cm, clip]{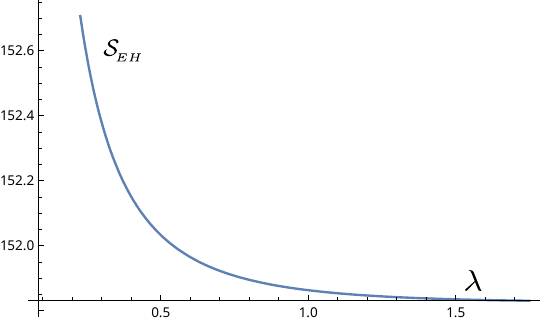}
\singlespace
\caption{Classical Einstein-Hilbert Action $\cS_{_{EH}}$ as function of $\l$ for fixed $\h$.\label{Fig:EHAction(lambda)}}
\vspace{-2mm}
\end{subfigure}
\hspace{1cm}
\begin{subfigure}{.44\textwidth}
\centering
\includegraphics[width=\linewidth,height=4.8cm, trim=0cm 1mm 0cm 0cm, clip]{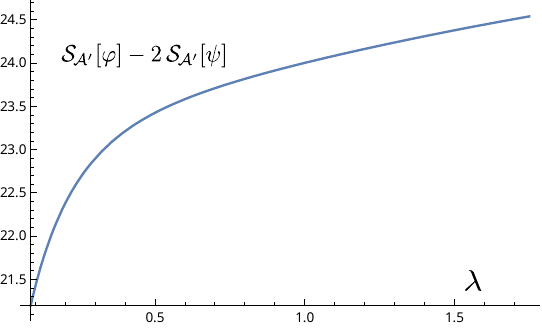}
\singlespace
\caption{Sum of Quantum Anomaly Actions as function of $\l$ for fixed $\h$. \label{Fig:AnomAction(lambda)}}
\vspace{-2mm}
\end{subfigure}
\caption{Contrasting behaviors of the classical action and quantum anomaly actions in (\ref{Sefflay}) as functions of $\l$ 
for the variational ansatz (\ref{ansfh}), (\ref{ansphipsi})-(\ref{Fansreg}), $\g=0.1$ and fixed $\l =0.231628$.  \label{Fig:Slam}}
\vspace{-3mm}
\end{figure}

The competition and balancing of the classical and higher derivative quantum terms is generic, as the existence of a minimum in 
Fig.~\ref{Fig:Min1} for $\g=1$ shows, even where the variational ansatz (\ref{ansfh}), (\ref{ansphipsi})-(\ref{Fansreg}) based on a 
small $\g$ approximation, cannot be expected to yield an accurate approximation.

\begin{figure}[h]
\includegraphics[width=8cm, height=5cm, trim=0cm 0cm 0cm 0cm, clip]{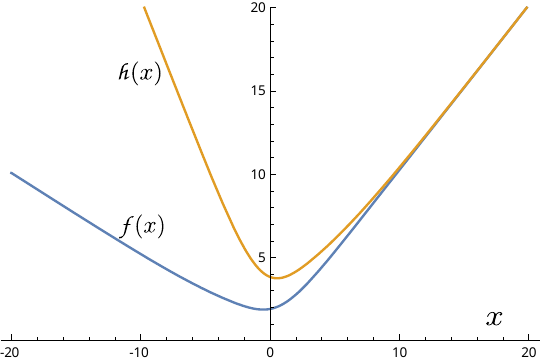}
\hspace{-1cm}
\singlespace
\caption{The rescaled metric functions $\mf(x)$ and $\mh(x)$ of (\ref{ansfh}) at the minimum of $\cS$ found in Fig.~\ref{Fig:Min0.1} 
for $\g=0.1$, showing that the sharp cusp of the classical solution (\ref{fhgrav}) illustrated in Fig.~\ref{Fig:SchwStar} 
is replaced by continuous behavior of the metric at $r=\rM, x=0$, due to the effect of the quantum anomaly terms.\label{Fig:Metric}}
\vspace{-3mm}
\end{figure}

Fig.~\ref{Fig:Metric} shows the rescaled metric functions $\mf, \mh$ for (\ref{ansfh}) at the values $(\h =2.55737,\l = 0.231628)$.
of the variational parameters at the minimum of the effective action of the boundary layer illustrated in Fig.~\ref{Fig:Min0.1} for $\g=0.1$,
due to the balance of classical and quantum anomaly terms. Since both $\mf$ and $\mh$ attain minima at $x \approx 0$ in the
central region of the boundary layer, but remain strictly positive throughout, the line element (\ref{sphsta}) with (\ref{mfmh})
describes an horizonless, static and non-singular compact object, a gravastar, rather than a black hole. Since the Killing norm 
$-K_\m K^\m = f >0$, there is no trapped surface and the Schwarzschild time $t$ is a globally defined timelike coordinate throughout.

The regularization of the geometry at $r=\rM$ removes the discontinuous step function behavior of the surface gravity of the
classical Schwarschild star of (\ref{surfgrav}), and the corresponding Dirac $\d$-function singularity of the Riemann curvature 
component $R^{tr}_{\ \ \,tr}$ of the classical solution in (\ref{dkapH}). Fig.~\ref{Fig:kappaH} shows that the surface gravity 
transitions continuously from its interior de Sitter value of $-0.5/\rM$ to its exterior Schwarzschild value of $+0.5/\rM$ through 
the central region of the boundary layer, described by the minimizing the effective action (\ref{Sefflay}) at a finite value of $(\h,\l)$. 
Correspondingly, since $dr = \e \rM dx = (\ell/2)\, dx$ by (\ref{epsdef})-(\ref{elldef}), Fig.~\ref{Fig:dkappaH} shows that
its derivative and $-\!\sqrt{\frac{f}{h}}\, R^{tr}_{\ \ \,tr}$ reach a finite maximal value of approximately $0.8/\ell^2$, in this central
region of the boundary layer, giving rise to a large transverse pressure from (\ref{Einstrans}), which integrates to the 
surface tension (\ref{tauS}). Although relatively large compared to the surrounding classical regions, this surface tension
and curvature is still far below the Planck scale, so that a semi-classical mean field treatment of the layer is admissible.

\begin{figure}
\vspace{-3mm}
\centering
\begin{subfigure}{.45\textwidth}
\includegraphics[width=\linewidth, height=5.2cm, trim=0cm 0cm 0cm 0cm, clip]{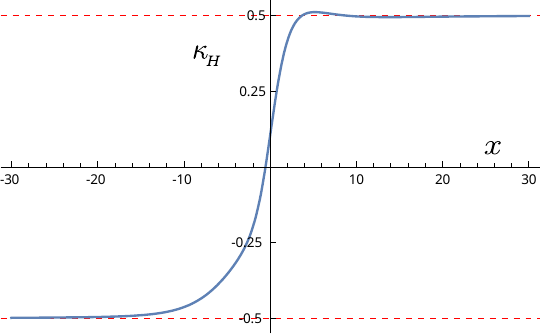}
\singlespace
\caption{$\ka_H$ in units of $1/\rM$ at the minimum of $\cS$. \label{Fig:kappaH}}
\vspace{-2mm}
\end{subfigure}
\hspace{1cm}
\begin{subfigure}{.44\textwidth}
\centering
\includegraphics[width=\linewidth,height=5.2cm, trim=0cm 1mm 0cm 0cm, clip]{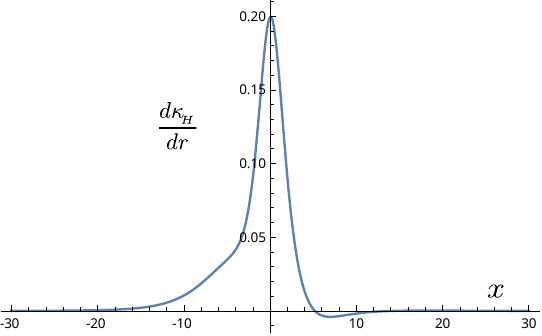}
\singlespace
\caption{$\frac{d \ka_H}{dr}$ in units of $1/\ell^2$ at the minimum of $\cS$. \label{Fig:dkappaH}}
\vspace{-2mm}
\end{subfigure}
\caption{The surface gravity $\ka_H$ of (\ref{surfgrav}) and its derivative related to the curvature component $R^{tr}_{\ \ \,tr}$
by (\ref{dkapH}), in the vicinity of $r=\rM,x=0$, for the minimum of $\cS$ found in Fig.~\ref{Fig:Min0.1} for $\g=0.1$. \label{Fig:kapdkap}}
\vspace{-3mm}
\end{figure}
\FloatBarrier

\section{Summary}
\label{Sec:Summary}

In this paper a non-singular solution for the final state of gravitational collapse consistent with quantum theory has
been proposed within the framework of the EFT of low energy gravity~\cite{EMEFT:2022}. 
The principal elements of this EFT and application to gravitational vacuum condensate stars are:
\begin{enumerate}[label= (\arabic*),leftmargin=8mm,topsep=1pt,itemsep=-1mm]
\item The effective action of the conformal anomaly (\ref{Sanom}), whose non-local macroscopic quantum correlations at 
lightlike separations are expressed by a long range, scalar conformalon field $\vf$, providing a relevant addition to the classical 
Einstein-Hilbert action of general relativity, and giving rise to large stress tensors on BH null horizons, 
{\it cf.} (\ref{ThorizM}), (\ref{ThorizH});
\item The replacement of the rigidly fixed cosmological constant $\La$ of classical general relativity by the 
action (\ref{Maxw}) of an exact $4$-form abelian gauge field $F= dA$, whose value and vacuum energy 
$\La_{\rm eff}$ (\ref{Lameff}) is dependent upon macroscopic boundary conditions rather than extreme UV scales;
\item The identification of $A$ with the Chern-Simons $3$-form of the Euler class defined in terms of the $S\!O(3,1)$
spin connection $\w^{ab}$ (\ref{eA})-(\ref{EdA}), which is {\it a priori} independent of the metric and Riemann curvature 
in a general Einstein-Cartan spacetime admitting torsion (\ref{Car2});
\item The extreme blueshifting of local energies (\ref{Tol}) in the near-horizon region allowing the lightest fermions of the SM,
minimally coupled to the spin connection $\w^{ab}$, to be treated as massless, contributing to the conformal anomaly,
and a $J\cdot A$ interaction (\ref{SanomF})-(\ref{Sint}) to the effective action;
\item The $3$-current source (\ref{Jphi}) for the `Maxwell' eq.~(\ref{Maxeq}) for $F$, allowing $F$ 
and $\La_{\rm eff}$ to change rapidly where $\vf$ does, in a thin boundary layer of skin depth of order the Compton wavelength 
(\ref{elltors}) of the lightest fermion of the SM, {\it i.e.}~$\sim 10^{-4}$ to $10^{-5}$ cm;
\item Activation of torsion (\ref{Car1}) and topological susceptibility $\big\lag \tF (x)\tF(x')\big \rag \propto \vk^4_{\rm eff}(x,x')$ 
{\it cf.}~(\ref{kapeffinv}) in this near-horizon region, producing a quantum vacuum phase boundary between regions of different 
vacuum energy, and realizing the quantum phase transition hypothesis of \cite{gravastar:2001,MazEMPNAS:2004,ChaplineHLS:2001};
\item A mean field solution of the metric, $F$ and the scalars $\vf,\j$ accounting for quantum correlations in this
boundary layer, by solving eqs.~(\ref{Feqloc})-(\ref{scEin}) resulting from the EFT action (\ref{Seff2});
\item The existence of the very small parameter $\e = L_{\rm Pl}/\rM \sim 10^{-38}$ of (\ref{epsdef}), characterizing the ratio 
of quantum to classical terms in the effective action (\ref{Seff2}), making the method of matched asymptotic solutions of boundary 
layer theory~\cite{BendOrzag} applicable; 
\item  A rescaled effective action (\ref{Sefflay}) and eqs.~(\ref{vfsoln})-(\ref{fheqs}) derived by rescaling of $r$ and the metric functions 
(\ref{xdef})-(\ref{mfmh}) in the boundary layer, retaining only the leading order in $\e \ll 1$.
\item The variational ansatz introduced in Sec.~\ref{Sec:BoundCond} indicating that a solution balancing the classical and quantum terms
in the action (\ref{Sefflay}) exists, in which the discontinuity $[\ka_H]$ in the surface gravity (\ref{surfgrav}) and resulting $\d$-function 
in the curvature (\ref{dkapH}) is replaced by a non-singular boundary layer with the same macroscopic surface tension,
with worldtube topology $\mathbb{R} \times \mathbb{S}^2$, as in Fig.~\ref{Fig:Tube}.
\end{enumerate}  

The finite skin depth boundary layer replaces both the classical BH horizon at $r\!=\!\rM$ and regularizes the infinitely sharp 
boundary of the classical constant density Schwarzschild star solution, which may be regarded as the classical limit of a gravastar, 
already inherent in classical general relativity from its inception.  The accurate numerical solution of the Euler-Lagrange eqs.~following 
from (\ref{Sefflay}) will be presented in a forthcoming paper. This solution, derived from the low energy EFT of gravity incorporating the 
the relevant macroscopic quantum effects in general relativity, will enable systematic study of the main properties of gravitational vacuum 
condensate stars, including their stability and normal modes of excitation, needed for predictions of observable gravitational wave 
signatures that could distinuish gravastars from classical BH's in present and future gravitational wave detectors.

\vspace{5mm}
\centerline{\bf Acknowledgements}

The author is indebted to Prof.~Pawel O. Mazur for innumerable insights on gravitational physics gained in discussions and 
collaborations over several decades, and to Evan David for his assistance in developing the Mathematica code used to produce
Figs.~\ref{Fig:Minima}-\ref{Fig:kapdkap} of Sec.~\ref{Sec:BoundCond}.

\vspace{3mm}


\appendix
\section{Curvature Components in Static, Spherically Symmetric Geometry}
\label{App:CurvStatSph}

The most rapid way to calculate all the curvature dependent quantities is by using the exterior calculus of differential 
forms. Defineing first the tetrad or orthonormal vierbein frame fields $e^a_{\ \m}$ in the cotangent space by the components 
of the one-forms
\vspace{-4mm}
\be
\mathsf{e}^a = e^a_{\ \m}\,dx^\m\qquad {\rm such\ that} \qquad e^a_{\ \m}\, e^b_{\ \n}\, \h_{ab} = g_{\m\n}
\vspace{-4mm}
\ee
where Greek indices are spacetime indices while Latin indices are tangent space indices,
$\h =$ diag (-+++) is the flat Minkowski metric, and $d$ here and below denotes the exterior derivative operator. 
From (\ref{sphsta}) we may take
\vspace{-4mm}
\begin{subequations}
\be
\begin{array}{llll}
\mathsf{e}^0 &= \sqrt{f} \, dt\,, \hspace{2cm}&e^0_{\ t} &= \sqrt{f}\\
\mathsf{e}^1 &= \displaystyle \frac{dr}{\!\!\sqrt h}\,, &e^1_{\ r} &= \displaystyle \frac{1}{\!\!\sqrt h}\\
\mathsf{e}^2 &= r\, d\th\,, &e^2_{\ \th} &= r\\
\mathsf{e}^3 &= r \sin\th\,d\f \,, & e^3_{\ \f} &= r \sin\th  
\end{array}
\vspace{-1mm}
\label{vierb}
\ee
\end{subequations}
with all other components that are not listed vanishing. The volume $4$-form is the wedge product
\vspace{-4mm}
\begin{align}
&\mathsf{e}^0 \wedge \mathsf{e}^1 \wedge \mathsf{e}^2 \wedge\mathsf{e}^3 =
\sdfrac{\!1}{4!}\, \ve_{\a\b\g\l} \,dx^\a \wedge dx^\b \wedge dx^\g \wedge dx^\l\nn
&= \sqrt{-g}  \, dt\wedge dr\wedge d\th\wedge d\f = \sqrt{\!\sdfrac{f}{h}}\, r^2 \sin\th \, dt\wedge dr\wedge d\th\wedge d\f
\label{vol4form}
\end{align}
\vspace{-1cm}

\noindent
where the spacetime Levi-Civita tensor is defined by
\vspace{-4mm}
\be
\ve_{\a\b\g\l} = \e_{abcd}\,  e^a_{\ \a}\,  e^b_{\ \b}\,  e^c_{\ \g}\,  e^d_{\ \l}
\label{Eps}
\vspace{-2mm}
\ee
with $\e_{abcd}$ the totally anti-symmetic symbol in tangent space and $\e_{0123} = +1$. Hence
\vspace{-4mm} 
\be
\ve_{tr\th\f} =  e^0_{\ t}\,  e^1_{\ r}\,  e^2_{\ \th}\,  e^3_{\ \f} = \sqrt{-g} = \sqrt{\!\sdfrac{f}{h}}\, r^2 \sin\th
\label{LCsphsta}
\vspace{-4mm}
\ee
in the coordinates (\ref{sphsta}). It is useful also to introduce the fully antisymmetric tensor density
\vspace{-2mm}
\be
[\a\b\g\l] \equiv \frac{1}{\!\!\sqrt{-g}}\, \ve_{\a\b\g\l} = \left\{
\begin{array}{cl}+1 & {\rm if}\ \a\b\g\l  \ {\rm is\ an\ even\ permutation\ of} \ tr\th\f\\
-1 & {\rm if}\ \a\b\g\l \ {\rm is\ an\ odd\ permutation\ of}\ tr\th\f\\
\ \,0 & {\rm if \ any\ two\ of\ the\ indices}\ \a\b\g\l \ {\rm are\ the\ same} 
\end{array}\right.
\label{epsdens}
\vspace{-1mm}
\ee
which takes on constant numerical values independent of the metric. Likewise
\vspace{-3.5mm} 
\be
\ve^{tr\th\f} = g^{tt} g^{rr} g^{\th\th} g^{\f\f} \!\sqrt{-g} = - \frac{\,1}{\!\!\sqrt{-g}}
\label{epsup}
\vspace{-3.5mm}
\ee
so that $\!\sqrt{-g}\, \ve^{\a\b\g\l}$ and hence (\ref{Sint}) are also independent of the metric. 

By definition the integral over the volume $4$-form (\ref{vol4form}) multiplied by any scalar function $\F(x) $ is
\vspace{-3mm}
\be
\int \F\ \mathsf{e}^0 \wedge \mathsf{e}^1 \wedge \mathsf{e}^2 \wedge \mathsf{e}^3 = \int d^4\!x \sqrt{-g}\ \F(x)  
=\! \int \! dt\,dr\,d\th\,d\f \sqrt{\sdfrac{f}{h}} \, r^2 \sin\th  \  \F\,.
\vspace{-2mm}
\ee
For spherically symmetric functions $\F(r)$ of $r$ only, the integral over the $\th, \f$ angular variables gives the total solid 
angle $4\p$, and with no time dependence, an overall factor of $\big[4 \p\! \int\! dt\big]$ can be eliminated from the integration 
measure and all terms in the effective action.

Referring to the definition of the torsion $2$-form (\ref{Car2}), and requiring it to vanish in Riemannian geometry 
gives the spin connection $1$-form $\w^{ab} = -\w^{ba}$ with the components
\vspace{-3mm}
\be
\begin{aligned}
\mathsf{\w}^{01} &= \frac{1}{2f}\frac{df}{dr} \sqrt{h}\, \mathsf{e}^0= \frac{1}{2}\frac{df}{dr} \sqrfr{h}{f}\,  dt\\
\mathsf{\w}^{02} &= \mathsf{\w}^{03} = 0 \\
\mathsf{\w}^{12} &= - \frac{\!\sqrt h}{r}\, \mathsf{e}^2 = -\sqrt{h} \,d\th\\
\mathsf{\w}^{13} &=  - \frac{\!\sqrt h}{r}\, \mathsf{e}^3 = -\sqrt{h}\, \sin\th\,d\f  \\
\mathsf{\w}^{23} &= -\frac{\cot \th}{r}\, \mathsf{e}^3 = -\cos\th\, d\f
\end{aligned}
\vspace{-3mm}
\ee
in terms of the metric functions in the coordinates of (\ref{sphsta}). From this the curvature $2$-form 
$\mathsf{R}^{ab} = -\mathsf{R}^{ba}$, defined by (\ref{Car2}) may be computed, yielding the non-vanishing components
\vspace{-2mm}
\begin{align}
\mathsf{R}^{01} &= -\frac{h}{4f^2} \left\{2f\, \frac{d^2 f}{dr^2}- \left(\frac{df}{dr}\right)^2  
+ \frac{f}{h}\,\left(\frac{df}{dr}\right)\left(\frac{dh}{dr}\right)\right\} \mathsf{e}^0\wedge \mathsf{e}^1 \nn[1ex]
\mathsf{R}^{02} &=- \frac{\,h}{2rf}\frac{df}{dr}\ \mathsf{e}^0\wedge \mathsf{e}^2\nn[1ex]
\mathsf{R}^{03} &=- \frac{\,h}{2rf}\frac{df}{dr}\ \mathsf{e}^0\wedge \mathsf{e}^3\\[1ex]
\mathsf{R}^{12} &=-\frac{1}{2r} \frac{dh}{dr}\ \mathsf{e}^1\wedge \mathsf{e}^2\nn[1ex]
\mathsf{R}^{13} &=-\frac{1}{2r} \frac{dh}{dr}\ \mathsf{e}^1\wedge \mathsf{e}^3\nn[1ex]
\mathsf{R}^{23} &=\frac{1-h}{r^2}\ \mathsf{e}^2\wedge \mathsf{e}^3\nonumber
\end{align}\\[-1cm]
so that there are just four independent non-vanishing components of the Riemann curvature tensor:
\vspace{-3mm}
\be
\begin{aligned}
\sA &\equiv R^{01}_{\ \ \, 01}= R^{tr}_{\ \ \, tr} = \displaystyle -\frac{h}{4} 
\left\{\frac{2}{f}\,\frac{d^2 \! f}{dr^2}- \frac{1}{f^2\!}\,\left(\frac{df}{dr}\right)^2 
+ \frac{1}{fh}\left(\frac{df}{dr}\right)\left(\frac{dh}{dr}\right)\right\}\\
&=  -\frac{1}{2}\sqrfr{h}{f}\frac{d}{dr} \left(\!\!\sqrfr{h}{f} \frac{df}{dr}\right)
= - \sqrfr{h}{f}\, \frac{d\ka_H}{\!dr}
\end{aligned}
\label{Rtrtr} 
\vspace{-3mm}
\ee
and
\vspace{-3mm}
\be
\begin{aligned}
\sB &\equiv R^{02}_{\ \ \, 02} =R^{03}_{\ \ \, 03} = R^{t\th}_{\ \ \, t\th}= R^{t\f}_{\ \ \, t\f}
=- \frac{\,h}{2rf}\frac{df}{dr}\\
\sC &\equiv R^{12}_{\ \ \, 12} =R^{13}_{\ \ \, 13} = R^{r\th}_{\ \ \, r\th}= R^{r\f}_{\ \ \, r\f}
= -\frac{1}{2r} \frac{dh}{dr}\\
&\hspace{8mm}\sD \equiv R^{23}_{\ \ \, 23} = R^{\th\f}_{\ \ \, \th\f} =\frac{1-h}{r^2}
\end{aligned}
\label{BCD}
\vspace{-1mm}
\ee
with all other components not listed or not related by symmetry vanishing identically. The $\sA = R^{tr}_{\ \ \,tr}$ 
component is the only one involving two derivatives of either of the metric functions $f$ or $h$, and the first form of (\ref{Rtrtr}) 
shows that it is generally singular at $f\!=\! -K^\m K_\n \!=\! 0$. The second form of (\ref{Rtrtr}) shows that a singularity appears
in $\!\sqrt{{f}/{h}}\, R^{tr}_{\ \ \,tr}$ if the surface gravity $\ka_H$ defined by (\ref{surfgrav}) is discontinuous at $f=0$, as occurs
in the constant density Schwarzschild star solution discussed in \cite{MazEM:2015} and Sec.~\ref{Sec:gravcl}.

From (\ref{Rtrtr}) and (\ref{BCD}) the non-vanishing components of the Ricci tensor
\vspace{-5mm}
\be
\begin{aligned}
R^t_{\ t} &= \sA +2\sB\\
R^r_{\ r} &= \sA+2\sC\\
R^\th_{\ \th} &= R^\f_{\ \f} = \sB + \sC+ \sD
\end{aligned}
\label{Ric}
\vspace{-2mm}
\ee
are obtained, as well as the Ricci scalar
\vspace{-4mm}
\be
R = 2\, \big(\sA+2\sB + 2\sC + \sD\big)
\label{Ricscal}
\vspace{-2mm}
\ee
and the Einstein tensor $G^\m_{\ \n} = R^\m_{\ \n} - R\, \d^\m_{\ \n}/2 \,$, 
\vspace{-4mm}
\be
\begin{aligned}
G^t_{\ t} &= -(2\sC+\sD) = \frac{1}{r} \frac{dh}{dr} - \frac{(1-h)}{r^2}\\
G^r_{\ r} &= -(2\sB +\sD)= \frac{h}{rf} \frac{df}{dr} - \frac{(1-h)}{r^2}
\end{aligned}
\label{Einstensor}
\ee
\vspace{-4mm}
\be
G^\th_{\ \th} = G^\f_{\ \f} = -(\sA +\sB + \sC) = 
\frac{1}{2}\sqrfr{h}{f} \frac{d}{dr} \left(\!\!\sqrfr{h}{f}\frac{df}{dr}\right)
+  \frac{h}{2rf} \frac{df}{dr} + \frac{1}{2r} \frac{dh}{dr} 
\label{Einstrans}
\ee
with all off-diagonal components vanishing. From (\ref{Einstensor}) one can check that the only non-trivial Bianchi identity
\vspace{-2mm}
\be
\na_\m G^\m_{\ r} =\frac{d}{dr}\, G^r_{\ r} + \frac{1}{2f} \frac{df}{dr}\, \Big(G^r_{\ r} - G^t_{\ t}\Big) 
+ \frac{2}{r} \left(G^r_{\ r} - G^\th_{\ \th}\right) = 0
\label{Gcons}
\vspace{-1mm}
\ee
is satisfied identically. The angular $G^\th_{\ \th}\!=\! G^\f_{\ \f}$ components are determined by symmetry and 
(\ref{Gcons}) in terms of the $G^t_{\ t}$ and $G^r_{\ r}$ components, which are the only two independent components needed.

Since from (\ref{Ricscal}),
\vspace{-5mm}
\be
r^2 \!\sqrfr{f}{h}\, R = - \frac{d}{dr} \left(r^2\! \sqrfr{h}{f} \frac{df}{dr}\right) + 2 \sqrfr{f}{h} \left(1 - h - r\,\frac{dh}{dr}\right)
\vspace{-2mm}
\ee
the classical Einstein-Hilbert action is
\vspace{-1mm}
\be
\displaystyle {S\!_{EH} =\frac{1}{16 \p G}\!\int\! d^4\!x\! \sqrt{-g}\, R =  \frac{\big[4\p \!\int\! dt\big]}{16 \p G} \!\int \! dr\,  r^2\! \sqrfr{f}{h}\, R 
= \frac{\big[4\p \!\int\! dt\big]}{8\p G} \!\int \! dr   \sqrfr{f}{h}\, \left(1-h  - r\,\frac{dh}{dr}\right) }
\label{EH}
\vspace{-2mm}
\ee
after discarding of the surface term, or equivalently, adding the  York boundary value term~\cite{York:1972} to $S\!_{EH}$.
Since the local variation
\vspace{-2mm}
\be
\d\! \int dr\! \sqrfr{f}{h}\,  r^2 R = 2\, \d \!\int dr \sqrfr{f}{h} \, \left(1-h  - r\,\frac{dh}{dr}\right)  
= - \sqrfr{f}{h}\, r^2 \,G^{\m\n}\, \d g_{\m\n} 
\vspace{-2mm}
\ee
the two independent components of the Einstein tensor $G^t_{\ t}$ and $G^r_{\ r}$ in (\ref{Einstensor}), are generated 
by variation of the action (\ref{EH}) with respect to $f(r)$ and $h(r)$ respectively. In general, 
\vspace{-2mm}
\be
 \d \!\int dr\! \sqrfr{f}{h}\,  r^2 \cL =\frac{1}{2} \sqrfr{f}{h}\, r^2 \,  T^{\m\n} \d g_{\m\n} 
=\frac{1}{2} \sqrfr{f}{h}\, r^2 \,\left( T^t_{\ t}\, \frac{\d f}{f} - T^r_{\ r}\, \frac{\d h}{h}\right)
\label{varfh}
\vspace{-3mm}
\ee
for any scalar Lagrangian $\cL$ containing only functions of $r$ in coordinates (\ref{sphsta}).

The Weyl conformal tensor
\vspace{-4mm}
\be
C^{\a\b}_{\ \ \,\g\l} = R^{\a\b}_{\ \ \,\g\l} - \sdfrac{1}{2} \left(\d^\a_{\ \g} R^\b_{\ \l} - \d^\a_{\ \l} R^\b_{\ \g} 
+ \d^\b_{\ \l} R^\a_{\ \g} - \d^\b_{\ \g} R^\a_{\ \l} \right) 
+ \sdfrac{1}{6} \left(\d^\a_{\ \g} \d^\b_{\ \l} - \d^\a_{\ \l} \d^\b_{\ \g}\right) R
\label{Weyldef}
\vspace{-4mm}
\ee
has the independent non-vanishing components 
\vspace{-4mm}
\begingroup
\allowdisplaybreaks[4]
\begin{align}
C^{tr}_{\ \ \, tr}  &= R^{tr}_{\ \ \, tr} -\sdfrac{1}{2}\left(R^t_{\ t} + R^r_{\ r}\right)  + \sdfrac{1}{6}\,R 
= \sdfrac{1}{3}\, \big(\sA-\sB-\sC-\sD\big)\nn
C^{t\th}_{\ \ \, t\th}= C^{t\f}_{\ \ \, t\f} &=R^{t\th}_{\ \ \, t\th} -\sdfrac{1}{2}\big(R^t_{\ t} + R^\th_{\ \th}\big)  + \sdfrac{1}{6}\,R
= -\sdfrac{1}{6}\,\big(\sA-\sB-\sC-\sD\big)\nn
C^{r\th}_{\ \ \, r\th}= C^{r\f}_{\ \ \, r\f}&= R^{r\th}_{\ \ \, r\th} -\sdfrac{1}{2}\big(R^r_{\ r} + R^\th_{\ \th}\big)  + \sdfrac{1}{6}\,R
= -\sdfrac{1}{6}\, \big(\sA-\sB-\sC-\sD\big)\nn
C^{\th\f}_{\ \ \, \th\f} &= R^{\th\f}_{\ \ \, \th\f} - R^\th_{\ \th}+ \sdfrac{1}{6}\,R= \sdfrac{1}{3}\, \big(\sA-\sB-\sC-\sD\big)
\label{Weyl}
\end{align}\\[-1.1cm]
\endgroup
all proportional to the same combination of $\sA-\sB-\sC+\sD$. Hence the Weyl tensor squared is
\vspace{-4mm}
\begin{align}
C^{\a\b\g\l}C_{\a\b\g\l} &= R^{\a\b\g\l}R_{\a\b\g\l} - 2R^{\a\b}R_{\a\b} + \sdfrac{1}{3} R^2
 = \sdfrac{4}{3}\,\big(\sA-\sB-\sC+\sD\big)^2\nn
&= \frac{1}{3} \left[-\sqrfr{h}{f} \frac{d}{dr} \left(\!\sqrfr{h}{f} \frac{df}{dr}\right) 
+ \frac{h}{rf} \frac{df}{dr} + \frac{1}{r} \frac{dh}{dr} + \frac{2\,(1-h)}{r^2}\right]^2
\label{Weylsq}
\end{align}\\[-1.1cm]
in the static, spherically symmetric geometry.

The $4$-form of the Euler class of (\ref{Fdef}) has the Hodge dual scalar
\vspace{-3mm}
\be
\begin{aligned} 
\ast \mathsf{F}&=  \sdfrac{1}{4}\ve_{\m\n\r\s} \e_{abcd}\, R^{ab\m\n}\, R^{cd\r\s}
=  \sdfrac{1}{4}\ve_{\m\n\r\s} \e_{\a\b\g\l}\, R^{\a\b\m\n}\, R^{\g\l\r\s}\nn
&\stackrel{Rie}{=} -\left\{R^{\a\b\g\l}R_{\a\b\g\l} - 4R^{\a\b}R_{\a\b} + R^2\right\} = - E
\label{Fstar}
\end{aligned}
\vspace{-3mm}
\ee
where the the notation $\stackrel{Rie}{=}$ denotes that the equality holds in a Riemannian geometry with zero torsion,
and $E$ is Riemannian Euler class invariant 
\vspace{-3mm}
\be
E= 8\, \big(\sA\sD + 2\sB\sC\big)= \frac{4}{r^2\!} \left[ \frac{h}{f} \frac{df}{dr} \frac{dh}{dr}
- (1-h) \sqrfr{h}{f} \frac{d}{dr} \left(\!\sqrfr{h}{f} \frac{df}{dr}\right)\right]
\label{EGB}
\vspace{-3mm}
\ee
in the coordinates of (\ref{sphsta}). This is a total covariant derivative, since
\vspace{-3mm}
\be
r^2 \!\sqrfr{f}{h}\,E = -4\, \frac{d}{dr} \left[(1-h)  \sqrfr{h}{f} \frac{df}{dr}\right]
\label{topE}
\vspace{-2mm}
\ee
is a total derivative and hence its integral is a topological quantity left invariant by any local variations that leave
its boundary conditions unchanged. This can be verified also from the fact that $\mathsf{F} = d\mathsf{A}$ is exact, 
with the $\mathsf{A}$ the Chern-Simons $3$-form (\ref{eA}), given explicitly by
\vspace{-4mm}
\begin{align}
\mathsf{A} &= 4\, \w^{01} \wedge d\w^{23} + 4\,\w^{23} \wedge d\w^{01} -8\, \w^{01}\wedge \w^{12} \wedge \w^{13}\nn
&= 2 \, \frac{d}{dr} \left(\!\!\sqrfr{h}{f} \frac{df}{dr}\right)\,\cos\th \,dt\wedge dr \wedge d\f 
+ 2\,(1-2h)\!\sqrfr{h}{f} \frac{df}{dr}\, \sin\th\, dt \wedge d\th \wedge d\f 
\label{3form}
\end{align}\\[-1cm]
in the coordinates (\ref{sphsta}), so that 
\vspace{-3mm}
\begin{align}
\mathsf{F} = d\mathsf{A} &= 
4\,\left[\! \sqrfr{h}{f} \frac{df}{dr} \frac{dh}{dr} - (1-h)\, \frac{d}{dr} \left(\!\sqrfr{h}{f} \frac{df}{dr}\right)\right]
\, \sin\th\, dt \wedge dr \wedge d\th \wedge d\f\nn
&= \frac{4}{r^2\!} \left[ \frac{h}{f} \frac{df}{dr} \frac{dh}{dr}
- (1-h) \sqrfr{h}{f} \frac{d}{dr} \left(\!\sqrfr{h}{f} \frac{df}{dr}\right)\right] \mathsf{e}^0 \wedge \mathsf{e}^1 \wedge \mathsf{e}^2 \wedge \mathsf{e}^3
\end{align}\\[-1cm]
which recovers (\ref{Fstar})-(\ref{topE}).

\section{Rescaling of Curvatures and Differential Operators in the Boundary Layer}
\label{App:Rescale}

Substituting the definitions (\ref{xdef})-(\ref{mfmh}) of the rescaled radial variable and metric functions in the boundary 
layer into the curvature terms of (\ref{Rtrtr})-(\ref{BCD}), one finds
\vspace{-4mm}
\be
\begin{aligned}
\sA &= -\frac{1}{2\e \rM^2}\sqrt{\frac{\mh}{\mf}}\frac{d}{dx} \left(\!\!\sqrt{\frac{\mh}{\mf}} \frac{d\mf}{dx}\right)  
\end{aligned}
\vspace{-3mm}
\label{Arescale}
\ee
is of order $1/\e$ and generically grows large in the near-horizon boundary layer if $f$ and $h$ are smooth differentiable
functions approaching zero as $r \to \rM$ according to (\ref{xdef})-(\ref{mfmh}). On the other hand
\vspace{-3mm}
\be
\sB  =- \frac{1}{2\rM^2 }\,\frac{\mh}{\mf}\frac{d\mf}{dx} + \cO (\e)\,, \qquad
\sC = -\frac{1}{2\rM^2} \frac{d\mh}{dx} + \cO (\e)\,,\qquad
\sD =\frac{1}{\rM^2} + \cO (\e)
\label{BCDrescale}
\vspace{-3mm}
\ee
are of order $\e^0$, and remain finite as $\e \to 0$. Thus $|\sB|,|\sC|,|\sD| \ll |\sA|$ and $\sB.\sC, \sD$ may 
be neglected relative to $\sA$ in the boundary layer.

This singular behavior as $\e \to 0$ of (\ref{Arescale}) is a regularized form of the Dirac $\d$-function (\ref{dkapH}), 
which appears if the metric functions $f$ and $h$ are only piecewise continuous, but fail to have continuous first 
derivatives at $r=\rM$, in the sense that the integral over a small interval around $r=\rM$
\vspace{-4mm}
\be
- \int_{\rM - \d r}^{\rM + \d r} dr \sqrt{\frac{f}{h}} \, R^{tr}_{\ \ tr}  \to  - \e \rM\!\!\int dx \sqrt{\frac{\mf}{\mh}} \,  
{\mbox {$\sA$}}= \frac{1}{2 \rM} \left[\!\sqrt{\frac{\mh}{\mf}} \frac{d\mf}{dx}\right]= [\ka_H] = \frac{1}{\rM}
\label{delkapH}
\vspace{-3mm}
\ee
reproduces the discontinuity in the surface gravity $\ka_H$ (\ref{dkapH}) of the classical Schwarzschild star (\ref{fhgrav}).
This finite difference of surface gravities $[\ka_H]$ follows from the boundary conditions (\ref{fbcs})-(\ref{hbc}) for $|x| \gg 1$ 
(but $\e |x| \ll 1$) in the rescaled boundary layer radial coordinate $x$. From (\ref{Einstrans}) and (\ref{Arescale})-(\ref{BCDrescale}) 
this same finite $[\ka_H]$ appears in $\sqrt{f/h}\, G^\th_{\ \th}$ to leading order in $1/\e$,  and hence Fig.~\ref{Fig:dkappaH}
implies that the transverse pressureattains a large but finite maximum value $p_\perp \simeq 0.8/(8\p G\ell^2)$ in the boundary layer. 

Retaining only the leading behaviors as $\e \to 0$ of (\ref{Ric})-(\ref{Ricscal}), and (\ref{boxPhi}) gives
\vspace{-5mm}
\be
\begin{aligned}
R^t_{\ t} \to R^r_{\ r}&\to \sA\\
R^\th_{\ \th} = R^\f_{\ \f} &= \sB + \sC + \sD = \cO(\e^0)\\[-.5ex]
R \to R^t_{\ t} + R^r_{\ r} \to 2\sA &=  \frac{1}{\e \rM^2}\, \cR
= -\frac{1}{\e \rM^2}\,\sqrt{\frac{\mh}{\mf}}\frac{d}{dx} \left(\!\!\sqrt{\frac{\mh}{\mf}} \frac{d\mf}{dx}\right)  
\end{aligned}
\label{RBoxlims}
\vspace{-3mm}
\ee
where the last term $2\sA$ is the scalar curvature of (\ref{sphsta}) with (\ref{xdef})-(\ref{mfmh}), and $dt = \rM d \bar t$, namely
\vspace{-3mm}
\be
ds^2 = \e \rM^2 \left(- \mf(x) \, d \bar t^{\,2} + \frac{dx^2}{\mh (x)} \right) + r^2  \,d \W^2 \to  
\e \rM^2 \left(- \mf(x)\, d \bar t^{\,2}+ \frac{dx^2}{\mh (x)} \right) = \e \rM^2 \,\big(d\bar s_2^2\big)
\label{2Dmetric}
\vspace{-3mm}
\ee
which reduces to the scalar curvature of the latter $2D$ metric to leading order in $\e$. Hence the geometry becomes effectively
two dimensional in the $\e \to 0$ limit, and $\cR$ is the dimensionless scalar curvature of the rescaled $2D$ line element $d\bar s_2^2$
of (\ref{2Dmetric}) in dimensionless ($\bar t, x$) coordinates near $r \simeq \rM$. In this limit the boundary layer 
at fixed $(t,\th,\f)$ becomes planar, and may be represented as in Fig.~\ref{Fig:Layer}.

From (\ref{Weyl})-(\ref{Weylsq}) and (\ref{topE}) we have  also
\vspace{-4mm}
\begin{align}
&C^{tr}_{\ \ \, tr}  \to \frac{1}{3}\, \sA = \frac{1}{\e \rM^2}\, \frac{\cR}{6}\nn
&C^{\a\b\g\l}C_{\a\b\g\l}\to \frac{4}{3}\, \sA^2 = \frac{1}{\e^2 \rM^4}\, \frac{\cR^2}{3} \nn
&E \to 8 \sA \sD = \cO(1/\e) 
\label{WeylErescale}
\end{align}\\[-1.2cm]
so that $E$ is of order $1/\e$ and can be neglected in comparison to $C^2$ to leading order in $\e$. 

A similar dimensional reduction to 2D occurs for differential operators operating on functions only 
of $r$ in the boundary layer. In the full 4D theory:
\vspace{-4mm}
\begin{align}
(\na \F)^2 &\equiv (\na^\a\F)(\na_\a\F) = h\, \left(\frac{d \F}{dr}\right)^2\nn
\na_t \na^t \F &= \frac{\,h}{2f}\frac{df}{dr} \frac{d \F}{dr}\nn[1ex]
\na_r \na^r \F &=  h\, \frac{d^2\F}{dr^2} +\frac{1}{2} \frac{d h}{dr} \frac{d \F}{dr}\nn[1ex]
\na_\th \na^\th \F &= \na_\f \na^\f\F = \frac{h}{r} \frac{d \F}{dr}
\label{vfderivs}
\end{align}\\[-1.2cm]
so that
\vspace{-1mm}
\be
\begin{aligned}
\sq \F &= \left(\na_t \na^t + \na_r \na^r +  \na_\th \na^\th + \na_\f \na^\f \right)  \F =
 h\, \frac{d^2\F}{dr^2} + \left(\frac{1}{2} \frac{d h}{dr} +  \frac{\,h}{2f}\frac{df}{dr} + \frac{2h}{r}\right)\frac{d \F}{dr}\\
&\hspace{2.5cm}=\frac{1}{r^2} \sqrfr{h}{f} \frac{d}{dr} \left(r^2 \!\sqrt{fh}\, \frac{d\F}{dr}\right) 
\label{boxPhi}
\end{aligned}
\ee
operating on any scalar function $\F = \F(r)$ in the static, spherically symmetric geometry of (\ref{sphsta}).

Substituting the definitions (\ref{xdef})-(\ref{mfmh}) of the rescaled radial variable and metric functions into (\ref{vfderivs})
shows that the $\na_t \na^t$ and $\na_r \na^r$ terms are of order $1/\e$ while the $\na_\th \na^\th, \na_\f \na^\f$ terms
are of order $\e^0$ and hence may be neglected in the boundary layer when $\e \ll 1$. Thus 
\vspace{-2mm}
\be
\begin{aligned}
&\sq \F \to  \big(\na_t\na^t + \na_r\na^r\big)\,\F  = 
\sqrfr{h}{f} \frac{d}{dr} \left(\!\sqrt{fh}\, \frac{d\F}{dr}\right) 
= \frac{1}{\e \rM^2} \sqrfr{\mh}{\mf} \frac{d}{dx} \left(\!\sqrt{\mf\mh}\, \frac{d\F}{dx}\right) 
= \frac{1}{\e \rM^2}\, \D_2 \F
\label{RBox2rescale}
\end{aligned}
\vspace{-2mm}
\ee
where $\D_2$, given by (\ref{Delta2}) is the wave operator in the dimensionless $2D$ metric $ds_2^2$ of (\ref{2Dmetric})
after rescaling by $\e\rM^2$. If the general $\F$ is replaced the scalar curvature $R$, we have also from (\ref{RBoxlims})
\vspace{-4mm}
\be
\sq R \to \frac{1}{\e^2 \rM^4}\, \D_2\cR
\label{sqRrescale}
\vspace{-3mm}
\ee
to leading $1/\e^2$ order in $\e$ for $\e \ll 1$ in the boundary layer.

Making use of (\ref{RBoxlims}) and (\ref{vfderivs})-(\ref{sqRrescale}), the operator $\D_4$ of (\ref{Deldef}) becomes
\vspace{-3mm}
\be
\D_4 \F \to  \frac{1}{\e^2 \rM^4} \,\left( \D_2^2 + \frac{\cR}{3} \D_2 + \frac{\mh}{3} \frac{d \cR}{dx}\frac{d}{dx}\right) \F
\equiv  \frac{1}{\e^2 \rM^4} \, \bar \D_4 \F
\label{Del4to2}
\vspace{-3mm}
\ee
in terms of rescaled 2D scalar curvature $\cR$ and wave operator $\D_2$, to leading order $1/\e^2$ order in $\e$ for $\e \ll 1$ 
in the boundary layer, for $\vf = \vf(r) \to \vf(x)$.

\section{The Anomaly Stress Tensor}
\label{App:TAnom}

The effective action of the conformal anomaly (\ref{Sanom}) $S\!_\cA = S\!_\cA^{\,(2)} + S\!_\cA^{\,(1)}$ contains terms both
quadratic and linear in $\vf$, denoted by superscripts$^{\,(2)}$ and$^{\,(1)}$ respectively, and gives rise to the stress-energy tensor 
\vspace{-3mm}
\be
T\!_{\cA'}^{\, \m\n} [\vf] \equiv \frac{2\!\!\!}{\!\!\sqrt{-g}}\, \frac {\d }{\d g_{\m\n}} \,S\!_{\cA'}[g;\vf] =
a'\, E^{\m\n}[\vf] + b\, C^{\m\n} [\vf]+ \sum_i \b_i \,T^{(i)\,\m\n}[\vf]
\label{Ttens}
\vspace{-1mm}
\ee
where $E^{\m\n}[\vf]$ is the metric variation of all terms proportional to $a$ in $S\!_\cA$ of (\ref{Sanom}), {\it viz.}~\cite{EMSGW:2017}
\vspace{-3mm}
\bea
&&E^{\m\n}[\vf] =  - 2\,(\na^{(\m}\vf) (\na^{\n)} \sq \vf)  + 2\na^\a \big[(\na_\a \vf)(\na^\m\na^\n\vf)\big]
- \tfrac{2}{3}\, \na^\m\na^\n\big[(\na_\a \vf)(\na^\a\vf)\big]\nn
&&\hspace{1.2cm}+\tfrac{2}{3}\,R^{\m\n}\, (\na_\a \vf)(\na^\a \vf) - 4\, R^{\a(\m}\left[(\na^{\n)} \vf) (\na_\a \vf)\right]
 + \tfrac{2}{3}\,R \,(\na^{(\m} \vf) (\na^{\n)} \vf) \nn
&&\hspace{.5cm} + \tfrac{1}{6}\, g^{\m\n}\left\{-3\, (\sq\vf)^2 + \sq \big[(\na_\a\vf)(\na^\a\vf)\big]
+ 2\, \big( 3R^{\a\b} - R g^{\a\b} \big) (\na_\a \vf)(\na_\b \vf)\right\}\nn
&&\hspace{-7mm}  - \tfrac{2}{3}\, \na^\m\na^\n \sq \vf  - 4\, C^{\m\a\n\b}\, \na_\a \na_\b \vf
- 4\, R^{\a(\m}\na^{\n)} \na_\a\vf  + \tfrac{8}{3}R^{\m\n}\, \sq \vf   +\tfrac {4}{3}\, R\, \na^\m\na^\n\vf
- \tfrac{2}{3} \left(\na^{(\m}R\right) (\na^{\n)}\vf)\nn
&& \hspace{1.8cm} + \tfrac{1}{3}\, g^{\m\n}\left\{ 2\sq^2 \vf + 6\,R^{\a\b} \,\na_\a\na_\b\vf
- 4 R \sq \vf  + (\na^\a R)(\na_\a\vf)\right\}\nn
&&\hspace{1cm}  = E^{(2)\,\m\n}[\vf] + E^{(1)\,\m\n}[\vf]
\label{Eab}
\eea
\vspace{-1.1cm}

\noindent 
containing terms both quadratic $E^{(2)\,\m\n}[\vf]$ and linear $E^{(1)\,\m\n}[\vf]$ in $\vf$, while
\vspace{-2mm}
\bes
\bea
&&C^{\m\n}[\vf] \equiv\ \frac{1\!}{\!\!\!\sqrt{-g}} \, \frac{\d}{\d g_{\m\n}} 
\left\{\! \int \!d^4 x \sqrt{-g}\,  C^2\,   \vf \right\} 
=\ -4\,\na_\a\na^\b\,\Big( C^{\a(\m\ \,\n)}_{\hspace{4mm}\b} \,\vf \Big)  -2\, C^{\a\m\ \,\n}_{\hspace{3mm}\b}\, R_\a^{\ \,\b}\, \vf\\
&& \hspace{2.5cm}T^{(i)\,\m\n}[\vf] \equiv  \frac{1\!}{\sqrt{-g}\ }\, \frac{\d}{\d g_{\m\n}}\left\{\int d^4\!x\sqrt{g}\,  \cL_i\, \vf\right\}
\eea
\label{CFterms}\ees
are the metric variations of the last two $b$ and $\b_i$  terms in (\ref{Sanom}) that are linear in $\vf$~\cite{EMSGW:2017,EMEFT:2022}.
The stress tensor appearing in (\ref{scEin}) resulting from the term quadratic in $\j$ is
\vspace{-3mm}
\be
T\!_{\cA'}^{(2)\, \m\n} [\j] = \frac{2\!\!\!}{\!\!\sqrt{-g}}\, \frac {\d }{\d g_{\m\n}} \left\{-2\, S\!_{\cA'}^{(2)}[g;\j]\right\} =
-2\, a'\, E^{(2)\,\m\n}[\vf] 
\label{T2psi}
\vspace{-3mm}
\ee
{\it i.e.} minus twice the quadratic term of (\ref{Eab}), with $\j$ substituted for $\vf$.

The rescaled stress tensor components obtained from (\ref{Eab}) defined by 
\vspace{-4mm}
\be
\begin{aligned}
-E^{t\,(2)}_{\ t}[\vf]& \to \frac{1}{\e^2 \rM^4}\,\vl_E^{(2)}[\vf]\,,\qquad&
-E^{t\,(1)}_{\ t}[\vf]& \to \frac{1}{\e^2 \rM^4}\,\vl_E^{(1)}[\vf] \\
E^{r\,(2)}_{\ r}[\vf]& \to \frac{1}{\e^2 \rM^4}\,\zp_E^{(2)}[\vf]\,,\qquad &
E^{r\,(1)}_{\ r}[\vf]& \to \frac{1}{\e^2 \rM^4}\,\zp_E^{(1)}[\vf]
\end{aligned}
\label{E12rescale}
\vspace{-3mm}
\ee
are
\vspace{-3mm}
\be
\begin{aligned}
\vl_E^{(2)}[\vf] &= (\D_2 \vf)^2 + 2\mh\,  \frac{d \vf}{dx} \left(\frac{d}{dx}\,\D_2 \vf\right) 
- \frac{1}{3} \left( 2 \mh \frac{d^2}{dx^2} + \frac{d \mh}{dx} \frac{d}{dx}\right) \left( \mh\,  \frac{d \vf}{dx} \frac{d \vf}{dx}\right)\\[1ex]
&\vl_E^{(1)}[\vf] = \frac{2}{3} \left\{ \cR\,  \D_2\vf +  \mh\, \frac{d \vf}{dx} \frac{d \cR}{dx}
+ \left( 2 \mh \frac{d^2}{dx^2} + \frac{d \mh}{dx} \frac{d}{dx}\right) (\D_2 \vf) \right\}
\end{aligned}
\label{rhoA}
\vspace{-1mm}
\ee
for the rescaled energy density terms, and 
\vspace{-4mm}
\begin{align}
\zp_E^{(2)}[\vf] &= -(\D_2 \vf)^2 + 2 \mh\,  \frac{d \vf}{dx} \left(\frac{d}{dx}\, \D_2 \vf\right) 
+ \frac{1}{3}\left( 2\, \cR + \frac{\mh}{\mf}\frac{d\mf}{dx}  \frac{d}{dx}\right)\left( \mh\,  \frac{d \vf}{dx} \frac{d \vf}{dx}\right)\nn 
&\zp_E^{(1)}[\vf] = \frac{2}{3} \left\{- \cR\,  \D_2\vf +  \mh\, \frac{d \vf}{dx} \frac{d \cR}{dx}
- \frac{\mh}{\mf}\frac{d\mf}{dx}\left(\frac{d}{dx}\D_2 \vf\right) \right\}
\label{pA}
\end{align}\\[-1cm]
for the corresponding rescaled pressure terms, appearing in the semi-classical Einstein eqs.~(\ref{scEin}) in the boundary layer.

For the anomaly stress tensor (\ref{Eab}), the quadratic and linear terms may be combined to define
\vspace{-2mm}
\be
\begin{aligned}
\vl_E[\vf] &\equiv \vl_E^{(2)}[\vf] + \vl_E^{(1)}[\vf] \\
\zp_E[\vf] &\equiv \zp_E^{(2)}[\vf] + \zp_E^{(1)}[\vf] 
\end{aligned}
\label{rhop12}
\vspace{-1mm}
\ee
for the total rescaled energy density and pressure resulting from $E^{\m\n}$ in (\ref{Eab}) appearing in the rescaled
semi-classical Einstein eqs.~(\ref{fheqs}) in the boundary layer.

The rescaled energy density and pressure arising from $C^{\m\n}$ in (\ref{CFterms}) which is linear in $\vf$, are defined similarly by
\vspace{-3mm}
\be
-C^t_{\ t}[\vf] \to \frac{1}{\e^2 \rM^2}\,\vl_C[\vf]\,,\qquad -C^r_{\ r}[\vf] \to \frac{1}{\e^2 \rM^2}\,\zp_C[\vf]\\
\label{Cabrescale}
\vspace{-3mm}
\ee
with
\vspace{-1mm}
\be
\begin{aligned}
&\vl_C[\vf] = \cR^2\,\vf + 2 \left( 2 \mh \frac{d^2}{dx^2} + \frac{d \mh}{dx} \frac{d}{dx}\right) (\cR \vf) \\[0.5ex]
&\zp_C[\vf] = -\cR^2\,\vf - 2 \, \frac{\mh}{\mf}\frac{d\mf}{dx}\frac{d}{dx} (\cR \vf)\,.
\end{aligned}
\label{rhopC}
\ee

\vfil
\pagebreak

\centerline{\large{\bf References}}
\bibliographystyle{apsrev4-1}
\vspace{-4mm}
\bibliography{gravity21Aug}

\end{document}